\documentclass[fleqn,usenatbib]{mnras}

\usepackage{newtxtext,newtxmath}
\usepackage{amsmath}
\usepackage{graphicx}	
\usepackage[fleqn]{nccmath}
\usepackage{xcolor}
\usepackage{multirow}

\usepackage[T1]{fontenc}

\DeclareRobustCommand{\VAN}[3]{#2}
\let\VANthebibliography\thebibliography
\def\thebibliography{\DeclareRobustCommand{\VAN}[3]{##3}\VANthebibliography}

\usepackage{graphicx}	
\usepackage{amsmath}	

\defcitealias{Shah2025a}{S2025LMC}

\renewcommand{\ne}{n_{\rm e}}

\newcommand{\cm}{\,{\rm cm}}    
\newcommand{\km}{\,{\rm km}}    
\newcommand{\pc}{\,{\rm pc}}     
\newcommand{\kpc}{\,{\rm kpc}}  
\newcommand{\sr}{\,{\rm sr}}     

\newcommand{\Msun}{\,{\rm M}_{\odot}} 

\newcommand{\Hz}{\,{\rm Hz}}    
\newcommand{\s}{\,{\rm s}}      
\newcommand{\kms}{\km\s^{-1}}    

\newcommand{\muG}{\,\mu{\rm G}} 

\newcommand{\Brms}{B_{\rm turb}}
\newcommand{\Bz}{\overline{B}_s}

\newcommand{\BznH}{\overline{B}_{\rm s, n_H}}
\newcommand{\Btrue}{B_{\rm true}}

\newcommand{\Bpredicted}{\overline{B}_{s,\rm pred}}

\newcommand{\Hi}{H\,\textsc{i}}
\newcommand{\Hii}{H\,\textsc{ii}}
\newcommand{\Hei}{He\,\textsc{i}}
\newcommand{\Heii}{He\,\textsc{ii}}
\newcommand{\Heiii}{He\,\textsc{iii}}
\newcommand{\nH}{n_{\rm H}}
\newcommand{\NH}{N_\mathrm{H\,\textsc{i}}}
\newcommand{\Tspin}{T_{\rm spin}}

\newcommand{\echi}{\sigma_{\chi}}

\newcommand{\LMCI}{\texttt{LMC-I}}
\newcommand{\LMCW}{\texttt{LMC-W}}
\newcommand{\MWI}{\texttt{MW-I}}

\newcommand{\erg}{\,{\rm erg}}  

\newcommand{\gizmo}{\textsc{gizmo}}
\newcommand{\grackle}{\textsc{grackle}}
\newcommand{\slug}{\textsc{slug}}

\usepackage{graphicx}	
\usepackage{amsmath}	
\usepackage{hyperref}
\usepackage{orcidlink}

\newcommand{\aref}[1]{\hyperref[#1]{Appendix~\ref{#1}}}

\title[Estimating B-fields]{
Accurate Extragalactic Magnetic Fields from Faraday Rotation with Optimal Dispersion Measure Estimators
}

\author[Shah et al.]{
Hilay Shah,$^{\orcidlink{0000-0002-9136-6731}, 1}$\thanks{E-mail: Hilay.Shah@anu.edu.au}
Mark R.~Krumholz,$^{\orcidlink{0000-0003-3893-854X},1}$,
N.\ M.\ McClure-Griffiths$^{\orcidlink{0000-0003-2730-957X},1,2}$, and
Zipeng Hu $^{\orcidlink{0000-0002-3758-552X}, 3}$
\\
$^{1}$Research School of Astronomy and Astrophysics, Australian National University, Canberra, ACT 2611, Australia\\
$^{2}$SKA Observatory, Jodrell Bank, Lower Whitington, Macclesfield, SK11 9FT, UK\\
$^{3}$Kavli Institute for Astronomy and Astrophysics, Peking University, Beijing 100871, China
\\
}

\date{Accepted XXX. Received YYY; in original form ZZZ}

\pubyear{\the\year{}}

\begin{document}
\label{firstpage}
\pagerange{\pageref{firstpage}--\pageref{lastpage}}
\maketitle

\begin{abstract}
Faraday rotation measures (RMs) are one of our few observational tools for measuring magnetic field strengths in extragalactic systems, but converting an RM to a magnetic field estimate requires knowledge of the electron column density -- the dispersion measure (DM) -- to the RM source. Because DMs are difficult to measure for most extragalactic radio galaxies, observers have adopted a range of strategies to estimate them from more easily measured quantities, but the accuracy of these approaches is poorly known. To address this, we carry out simulated observations of high-resolution magnetohydrodynamic simulations of a range of galactic environments to explore the performance of various possible DM estimators. We obtain the best results using an estimator $\mathrm{DM} \propto \mathrm{EM}^{\alpha} \ N_\mathrm{H\,\textsc{i}}^{\beta}$, where EM is the emission measure and $N_\mathrm{H\,\textsc{i}}$ is the atomic hydrogen column density, with exponents $\alpha \approx 0.2-0.4$ and $\beta \approx 0-0.1$ depending on galactic environment (e.g., galaxy centres versus outskirts, and dwarfs versus spirals). We show that the variation of these exponents with environment can be understood in terms of simple physical arguments. Based on our tests, we provide recommended best practices for extracting galactic magnetic fields from RM data as a function of galactic environment and of proxy data availability, and show that using these methods one can obtain field measurements that are accurate to a few tenths of a dex. This work therefore represents an important step toward making use of RM data from next-generation surveys with the SKA.
\end{abstract}




\section{Introduction}
\label{sec:intro}

Magnetic fields are found at scales ranging from planetary ($\sim 10^{-10}-10^{-8} \pc$) to intergalactic ($\sim 10^6 \pc$). On galactic scales, they potentially affect a vast range of astrophysical processes, from galaxy mergers \citep[e.g.,][]{Whittingham2023} to formation of circumnuclear structures \citep[e.g.,][]{Moon2023} to galactic wind propagation \citep[e.g.,][]{Das2023} to mixing and energy balance in the circumgalactic \citep{Shah2025b} and interstellar \citep{Seta2025} media to star formation \citep{Krumholz19b}. Despite their importance, magnetic fields are notoriously difficult to measure, which limits our ability to place observational constraints on their importance. 

There are four major techniques to trace magnetic fields \citep{Klein2015, ShukurovSubramanian2021}: (i) linear polarisation of synchrotron emission, (ii) Faraday rotation of linearly-polarised emission passing through a magnetised plasma, (iii) Zeeman splitting of atomic and molecular spectral lines, and (iv) polarised infrared emission or optical absorption from aspherical dust grains whose axes align with magnetic fields.

While all of these techniques can provide extrgalactic measurements to some extent, only (i), (ii), and (iv) can consistently map galactic-scale magnetic fields in extragalactic sources. Out of these, (i) and (iv) rely on regions from which either synchrotron emission or polarised dust emission is produced \citep{Fletcher2011, Beck2015, Pattle2021, Lopez-Rodriguez2022, Clements2025}. Moreover, they are vulnerable to depolarisation both by foreground structures that destroy the polarised signal farther from the observer \citep{Mao2012, Kierdorf2020} and by averaging over a telescope beam that may, for a distant source, smear together numerous field reversals that destroy the signal. Consequently, technique (ii) -- rotation measures (RMs) -- is often preferred. Due to the pencil beam nature of the background sources used in this method, beam and foreground depolarisation are much less of a concern. In addition, the method provides the value of the magnetic field integrated over the entire line of sight rather than localised to a potentially small and unrepresentative emitting region.

A number of studies using this method have been carried out \citep[e.g.,][]{Gaensler2005, McClure-Griffiths2010, Hill2013, Betti2019, LanP2020, AmaralEA2021, Shah2021, Jung2021, Heesen2023, Livingston2024, Carretti2025}, and have provided robust maps of magnetic fields for different extragalactic sources at various redshifts and scales. The main drawback of the RM method is that it relies on the presence of sufficiently bright linearly polarised background sources -- usually radio quasars -- and the density of suitable sources limits the angular resolution of the measurements. Increasing telescope sensitivity can therefore increase the effective angular resolution of RM measurements, by allowing use of fainter background sources with a higher density on the sky; doing so is one of the key science goals of some of the biggest radio telescopes in the world, including the Low Frequency Array (LOFAR; \citealt{Osullivan2023}), Australian SKA Pathfinder (ASKAP; \citealt{Johnston2008, Thomson2023, Gaensler2025}), and the future Square Kilometre Array (SKA; \citealt{Gaensler2004, Gaensler2010, Thomson2023}). A first step in this campaign, the Spectra and Polarisation in Cutouts of Extragalactic sources from RACS (SPICE-RACS) survey \citep{Thomson2026}, recently reported detection of $\approx 3 \times 10^5$ RM sources across the sky, with an areal density of $\approx 6.7$ per $\rm deg^2$, significantly higher than previous efforts. The ongoing POSSUM survey \citep{Gaensler2025} will improve this by a factor of nearly 7. Upcoming surveys from the SKA will map the RM grid at even higher resolutions \citep{Beck2004, Heald2020}.

To deliver the maximum scientific benefit from these radio telescopes, however, it is of the utmost importance to retrieve magnetic fields from the RM measurements with the highest possible accuracy -- and this is not a small challenge. RM is the path integral of the electron density multiplied by the magnetic field along the line-of-sight (LOS) to the emission source. To accurately retrieve magnetic fields, one must therefore break the degeneracy between electron density and magnetic field strength. This requires an independent measurement of the electron column density along the path to the source, known as the dispersion measure (DM). DMs are only directly measurable for pulsed rather than steady sources, and thus are typically obtained from pulsar or Fast Radio Burst (FRB) observations \citep{Manchester1972, Lorimer2007}. However, only a relatively small number of extragalactic pulsars are known, and almost none beyond the Magellanic Clouds. The lack of pulsar observations and the associated DM measurements forces us to estimate the DM from other, more easily observed quantities in extragalactic sources. 

Popular proxies for DMs in extragalactic sources include emission measures (EMs) obtained from either optical recombinations or radio free-free emission, and neutral hydrogen column densities ($\NH$) \citep{Heiles1980, Heiles1981, Mao2008, Harvey-Smith2011, Kaczmarek2017, Livingston2022, Livingston2024}. In some cases authors forgo the use of proxies entirely, and simply adopt values derived from Milky Way pulsar observations  \citep[e.g.,][]{Shah2021} or taken directly from simulations \citep{Carretti2025}. These models, in turn, rely on differing assumptions about the physical structure of the ISM being probed, and the accuracy of these assumptions is unknown. There is significant cause for concern, since the individual approaches often yield inconsistent results -- $\gtrsim 100\%$ disagreement in some cases -- for the magnetic field strength from the same underlying RM data (see \autoref{ssec:comparison} for more details).

In this study, we seek to improve this situation by using high-resolution simulations of a spiral galaxy, an isolated dwarf, and an interacting dwarf to provide the first estimates of the accuracy (or lack thereof) of various assumptions commonly used in the literature to estimate DMs. We quantify the errors and biases of the most common approaches, but most importantly, we also provide ``best practice'' approaches to predict magnetic fields, calibrated based on our numerical models, for an observer probing different galaxies in varying environments. 

The paper is organised as follows. In \autoref{sec:methods}, we outline our simulation setup, post-processing and radiative transfer pipeline for the generation of the mock observables and physical quantities that an observer would derive from the simulated galaxies. \autoref{sec:results} outlines our results and presents measurements for how well DMs correlate with various possible proxies for them in different galactic environments. \autoref{sec:discussion} provides a physics-based understanding of the magnetic field prediction models detailed in \autoref{sec:results}, and we bring this understanding together in \autoref{sec:to_observers} to provide our final recommendations to observers. Finally, in \autoref{sec:conclusions} we summarise our results.

\section{Methods}
\label{sec:methods}
Here we first introduce the simulations we analyse in \autoref{sec:Simulation}, and then in \autoref{sec:MockObservables} discuss how we generate several mock observables using radiative transfer.

\subsection{Simulations}
\label{sec:Simulation} 
In this paper, we analyse three MHD simulations, using the final time snapshot from each: an isolated dwarf galaxy modelled on the LMC, the same LMC-like galaxy but in an environment where it undergoes ram pressure stripping modelled on that to which the actual LMC is subject, and an isolated Milky Way (MW)-like galaxy. \citet[hereafter \citetalias{Shah2025a}]{Shah2025a} describe the simulation setup and initial conditions for the two LMC-like cases, while the Milky Way-like case was initially described by \citet{Wibking2023}, then re-simulated for a shorter period at higher resolution by \citet{Hu2023} and \citet{Hu2024}; we use the higher-resolution re-simulation in this work. For brevity, in the remainder of this paper, we assign shorthand labels to our three simulations: \LMCI\ and \LMCW\ for the isolated LMC and LMC subject to a wind (these correspond to the runs labelled \texttt{LMC-I-2-6-M} and \texttt{LMC-W-2-6-M} in \citetalias{Shah2025a}), and \MWI\ for the Milky Way-like galaxy. Below we briefly describe the setup we use for all simulations (\autoref{sec:Setup}) and the differences in the initial conditions between the three cases we use (\autoref{sec:InitialConditions}). These summaries are provided for convenience, and we refer readers to the original papers describing these simulations for full details. In \aref{sec:robustness_tests} we also repeat our analysis for some of the additional simulations presented in \citetalias{Shah2025a}, which use different resolutions or slightly different initial conditions, and we verify that our major qualitative conclusions hold equally well for these runs. For brevity we therefore focus only on \LMCI, \LMCW, and \MWI~in the main text.

\subsubsection{Setup}
\label{sec:Setup}
All three simulations were run using the \gizmo~code \citep{Hopkins2015std} with ideal-magnetohydrodynamics (MHD) and self-gravity in the meshless finite mass (MFM) setting \citep{Hopkins2015mhd} with constrained-gradient divergence cleaning \citep{Hopkins2016divb}. Our chemistry network uses the \textsc{Grackle} chemistry and radiative cooling library \citep{Smith2008, Smith2016}, which allows us to track the time-dependent abundances of \Hi, \Hii, \Hei, \Heii, \Heiii, and free electrons, and to incorporate gas cooling using tabulated H, He, and metal cooling rates calculated with the photo-ionisation code \textsc{Cloudy} \citep{Ferland1998, Cloudy23}. The gas is also subject to radiative heating, which is provided by a UV radiation field modelled after \citet{HM2012} at redshift 0 and an FUV background that produces a uniform photoelectric heating rate $8.5 \times 10^{-26} \nH \erg\ \s^{-1}$, where $\nH$ is the number density of H nucleons, in gas with temperatures $T < 2\times 10^4$ K \citep{WolfireEA2003, Tasker&Bryan2008}.

The simulation handles star formation and feedback by stochastically converting gas particles into collisionless star particles when their densities $\rho$ exceed a threshold value $\rho_\mathrm{SF}$, where $\rho_\mathrm{SF}$ is chosen so the simulation mass resolution $\Delta m$ equals the Jeans mass evaluated at $\rho_\mathrm{SF}$ and the corresponding equilibrium temperature $T_\mathrm{eq,SF}$; i.e., $\Delta m = M_J(\rho_\mathrm{SF}, T_\mathrm{eq,SF})$. The probability that a gas particle is converted to stars is set so that the expected mass converted to stars per free-fall time $\epsilon_{\mathrm{ff}} = 0.01$, consistent with observations \citep{Krumholz2019}; in order to avoid infeasibly small time steps, this is increased greatly for densities $\rho > 100\rho_\mathrm{SF}$ in order to force rapid conversion to stars in unresolvably dense gas.

Once stars form, they provide feedback via three pathways: stellar winds, supernovae, and photoionisation. Stellar winds and supernovae (SN) use an explicit mechanical feedback coupling algorithm that conserves momentum and energy, and interpolates SN feedback between thermal energy or mechanical momentum deposition depending on whether the Sedov-Taylor radius is resolved \citep{Hopkins2014feedback, Hopkins2018feedback, Wibking2023}. For photoionisation we use a Str\"omgren volume approach to heat the gas within the Str\"omgren sphere around each star particle to $10^4$ K (\citealt{Armillotta2019}, based on \citealt{Hopkins2018feedback}). Our simulation achieves a much higher mass resolution than cosmological simulations, necessitating a stochastic star-by-star treatment as outlined in \citet{Hu2023}. Formation of each star particle of mass $M_*$ triggers a stochastic draw from a stellar population of that mass from a Chabrier IMF \citep{Chabrier2005} using the \slug~(Stochastically Lighting Up Galaxies) stellar population synthesis code \citep{daSilva2012, Krumholz2015}. \slug, running as backend, tracks stellar population evolution with age and returns the ionising luminosity, wind luminosity, and timing of all supernovae (with an assumed energy release of $10^{51}$ erg per supernova).

On top of the on-the-fly implementation of physics described above, we postprocess some quantities to account for the fact the the assumption of a uniform background radiation field, while reasonable for on-the-fly simulation because it has little impact on the dynamics, does not provide a realistic ionisation structure for the denser phases of the ISM (see Section 2.4 of \citetalias{Shah2025a} for more details). In reality, the UV background radiation field should shift from an intergalactic one like the \citet{HM2012} background to a spectral shape characteristic of the interstellar radiation field (ISRF) as one moves from outside to inside the Reynolds layer in the ISM \citep{Reynolds1993}. This change in turn requires re-running the Str\"omgren volume calculation to generate local photoionisation structure consistent with the altered background radiation field. To this end, we use \textsc{Cloudy} to generate tables for equilibrium ionisation states for all species of interest with a background radiation field following \citet{Black1987}, subject to extinction from a neutral hydrogen slab with column density $10^{21} \cm^{-2}$ (typical for the ISM -- \citealt{Stil2002, Draine2011, Saha2018}), and a cosmic ray ionisation rate of $4.6 \times 10^{-16} \s^{-1}$ \citep{Indrirolo2007}. We replace the \grackle-predicted ionisation state with this new calculation in gas cooler than $5000$ K, a value chosen by calibrating the simulations against observations of pulsar dispersion measures in the LMC -- see \citetalias{Shah2025a} for details. We then re-run the Str\"omgren volume calculation to re-impose the effects of local photoionisation. While in \citetalias{Shah2025a} we applied this procedure only to \LMCI~and \LMCW, for consistency in this paper, we apply the same procedure to \MWI.

\subsubsection{Initial Conditions}
\label{sec:InitialConditions}

\begin{table*}
\centering
\caption{Simulation parameters}
\begin{tabular}{lllrrr}
\hline\hline
\\[-2ex]
Parameter & Unit & Meaning & \multicolumn{3}{c}{Simulation} \\
& & &
\LMCI & \LMCW & \MWI
\\[0.1ex]
\hline
\\[-2ex]
$\Delta m_\mathrm{g}$ & M$_\odot$ & Gas mass resolution & 250 & 250 & 89 \\
$\Delta m_\mathrm{DM}$ & $10^4$ M$_\odot$ & Dark matter mass resolution & 2.017 & 2.017 & 12.54 \\
$\Delta m_*$ & M$_\odot$ & Old stellar particle mass resolution & 284 & 284 & 3437 \\
$\rho_\mathrm{SF}/m_\mathrm{H}$ & cm$^{-3}$ & Star formation threshold normalised to hydrogen mass & 1018 & 1018 & 1000 \\
$R_0$ & kpc & Gas disc scale length & $4.8$ & $4.8$ & $3.43$ \\
$Z_0$ & kpc & Gas disc scale height & $0.48$ & $0.48$ & $0.343$\\
$R_*$ & kpc & Stellar disc scale length & $1.8$ & $1.8$ & - \\
$Z_*$ & kpc & Stellar disc scale height & $0.3$ & $0.3$ & $0.35$\\
$B_0$ & $\muG$ & Ordered magnetic field strength & $2$ & $2$ & $10$\\
$\Brms$ & $\muG$& Turbulent magnetic field RMS strength & $6$ & $6$ & $0$\\
$M_{200}$ & $10^{12}$ $\Msun$ & Dark matter halo mass & $0.1775$ & $0.1775$ & $1.07$\\
$M_\mathrm{*,d}$ & $10^9$ $\Msun$ & Stellar disc mass & $2.5$ & $2.5$ & $34$\\
$M_\mathrm{g,d}$ & $10^9$ $\Msun$ & Gas disc mass & $2.2$ & $2.2$ & $8.6$\\
$M_\mathrm{*,b}$ & $10^9$ $\Msun$ & Stellar bulge mass & - & - & $4.3$
\\ \hline\hline
\end{tabular}
\label{tab:InitialConditions}
\end{table*}

We collect all the parameters describing these simulations in \autoref{tab:InitialConditions}. The first three of these describe the mass resolution for the four different particle types included in the simulations: gas particles with mass resolution $\Delta m_\mathrm{g}$, dark matter with mass resolution $\Delta m_\mathrm{DM}$, old stellar particles with mass resolution $\Delta m_*$, and young stellar particles which have the same mass $\Delta m_\mathrm{g}$ as the gas particles from which they form. The old stellar particles do not have a \textsc{slug} model attached to them, while the young stellar particles do and therefore provide feedback. Only the first three particle types are present in the initial conditions.

For the two LMC-like simulations, the dark matter and old stellar particle distributions are initialised to an $N$-body equilibrium generated by the \textsc{GALIC} code \citep{Yurin2014}, while for the MW case \citet{Wibking2023} take the corresponding initial distributions from the equilibrium initial conditions adopted from the \textsc{AGORA} isolated disc galaxy setup \citep{Kim2016}. In all these setups, the stellar disc is a double-exponential with radial and vertical scale lengths $R_*$ and $Z_*$ and total mass $M_\mathrm{*,d}$, while the dark matter halo is spherical and has total mass $M_{200}$; \MWI\ also includes an initial stellar bulge with total mass $M_\mathrm{*,b}$.

All simulations also include an initial gas disc of mass $M_\mathrm{g,d}$; to initialise this component of the galaxy we sample article positions from double-exponential distribution $p(r,\phi,z) \propto r \exp \left(-r/R_0\right) \exp\left(-|z|/Z_0\right)$, where $(r,\phi,z)$ are the standard cylindrical coordinates. The gas disc particle velocities follow a centrifugal equilibrium, generated using the \textsc{pytreegrav} package\footnote{\url{https://github.com/mikegrudic/pytreegrav}} for the LMC, and \citet{Springel2005}'s implementation for the MW. We initialise gas magnetic fields with a combination of ordered azimuthal component, given by $B_{\phi} = B_0 \exp\left(-r/R_0\right) \exp\left(-|z|/Z_0\right)$, and a turbulent component, represented by a divergence-free Gaussian random field with a Kolmogorov power spectrum (slope $-5/3$; \citealt{Minter1996}), standard deviation $\Brms$, and spatial scales ranging from the resolution limit to 1 $\kpc$. The initial divergence cleaning \citep{Hopkins2016divb} gets rid of any numerical divergence present in the initial conditions. Finally, the \LMCW\ simulation contains, in addition to the particles making up the galaxy, include a set of gas particles representing the Milky Way CGM through which the LMC is falling; we refer readers to \citetalias{Shah2025a} for a full description of how that background and the LMC falling through it are initialised.

\subsection{Mock Observables}
\label{sec:MockObservables}

In this section, we describe how we calculate the mock observables that will be the focus of this work. We first define the observables of interest in \autoref{ssec:definitions}, introduce the quantities derived from them in \autoref{ssec:derived_quantities}, and finally describe how we generate samples of sightlines on which to carry out our analysis in \autoref{sssec:samples}.

\subsubsection{Definitions of the observables}
\label{ssec:definitions}

We begin by defining the quantities on which the remainder of our analysis relies: rotation measure (RM), dispersion measure (DM), emission measure (EM), velocity-integrated H$\alpha$ intensity, free-free intensity, and \Hi-21 cm spectrum. Consider a line of sight (LOS) through the simulation, and let $s$ define position along the LOS, with $s = 0$ corresponding to the position of the observer. The most straightforward and unambiguous quantities to define are RM, DM, EM, and true \Hi~column density; we define these as
\begin{equation}
    \left(\begin{array}{c}
    \mathrm{RM} \\
    \mathrm{DM} \\
    \mathrm{EM} \\
    N_\mathrm{\Hi,true}
    \end{array}
    \right)
    = 
    \int_0^\infty \left(
    \begin{array}{c}
    k_\mathrm{RM} n_e (-B_s) \\
    n_e \\
    n_e^2 \\
    n_\mathrm{H^0}
    \end{array}
    \right) ds,
\end{equation}
where $n_e$ is the electron number density, $n_\mathrm{H^0}$ is the number density of free atomic, neutral hydrogen, $B_s$ is the component of the magnetic field along the LOS, $k_\mathrm{RM} = e^3 / 2\pi m_e^2 c^4 = 0.812$ rad m$^{-2}$ $(\mathrm{pc}\, \mathrm{cm}^{-3})^{-1}$ is the rotation measure constant, and the integrals are computed along the LOS. Note the negative sign appearing before the $B_s$ term ensures that RM is positive (negative) when magnetic fields point towards (away from) the observer, which is the standard convention followed by radio astronomers \citep{Ferriere2021}. We defer a discussion of how we evaluate these and the other integrals appearing in this section numerically to \aref{sec:AdaptiveBinning}.

For the velocity-integrated H$\alpha$ and free-free surface intensities, we require models for the emissivity of H$\alpha$ and free-free. For these we adopt approximations from \citet{Draine2011}, which give
\begin{equation}
    \left(\begin{array}{c}
    W_{\mathrm{H}\alpha} \\
    I_{\nu,\mathrm{ff}}
    \end{array}
    \right)
    = 
    \int_0^\infty \left(
    \begin{array}{c}
    \alpha_{\mathrm{H\alpha,eff}} h \nu_\mathrm{H\alpha} n_e n_\mathrm{H^+} / 4\pi \\
    \mathcal{L}_{\nu,\mathrm{ff}} n_e n_\mathrm{H^+}
    \end{array}
    \right) ds,
\end{equation}
where $n_\mathrm{H^+}$ is the proton number density, $\nu_\mathrm{H\alpha}=457$ THz is the frequency of the H$\alpha$ line,  and the approximate expressions we adopt for the coefficients appearing in this expression are
\begin{eqnarray}
    \label{eq:alpha_eff}
    \alpha_\mathrm{eff,H\alpha} & \approx & 1.17 \times 10^{-13} T_4^{-0.942-0.03\ln T_4} \;\mathrm{cm}^3\;\mathrm{s}^{-1} \\
    \mathcal{L}_{\nu,\mathrm{ff}} = \frac{j_{\nu,\mathrm{ff}}}{n_e n_\mathrm{H^+}} & \approx & 3.35\times 10^{-40} \nu_9^{-0.118} T_4^{-0.323}\;\mathrm{erg}\;\mathrm{cm}^3\;\mathrm{s}^{-1}\;\mathrm{Hz}^{-1}.
\end{eqnarray}
Here $T_4$ is the gas temperature normalised to $10^4$ K, and $\nu_9$ is the observing frequency normalised to $10^9$ Hz. In practice we compute free-free emission at 1 GHz for the remainder of this work. We pause to note a few caveats with regard to these expressions. First, note that $I_{\nu,\mathrm{ff}}$ and $W_\mathrm{H\alpha}$ do not have the same units, because $I_{\nu,\mathrm{ff}}$ is an intensity but $W_\mathrm{H\alpha}$ is a frequency-integrated intensity, since it represents the total emission in the H$\alpha$ line integrated over all velocities. This difference simply reflects the fact that observations of free-free emission are always spectrally resolved, while most observations of the H$\alpha$ line are not. Second, for free-free, we are ignoring the subdominant contribution from helium. Since this is only a $\approx 10\%$ effect, the error we make by doing so is small. Third, our expressions here ignore absorption. For free-free, this choice is reasonable, since none of our simulated galaxies have ionised gas densities high enough that they are likely to become optically thick at 1 GHz. For H$\alpha$, neglect of absorption is more questionable, since dust opacities are likely non-negligible, but we assume for simplicity that the effects of dust absorption can be removed by standard methods such as the Balmer decrement; properly modelling the effects of dust absorption and then the observational techniques used to remove those effects is beyond the scope of this paper.

Finally, for the 21 cm line, we consider a LOS passing through the simulation domain to a background source with intensity $I_0$, and then compute the effects of both emission and absorption from within the volume of the simulation. The observed intensity at frequency $\nu$ is given by the usual formal solution to the equation of radiative transfer,
\begin{equation}
    I_\nu = I_0 e^{-\tau_\nu} + \int_0^{\tau_\nu} e^{-(\tau_\nu - \tau')} S_\nu \, d\tau',
    \label{eq:Inu}
\end{equation}
where $\tau_\nu = \int_0^\infty \kappa_\nu \, ds$ is the optical depth, $S_\nu = j_\nu/\kappa_\nu$ is the source function, $\tau'$ is the optical depth between the source and the emitting region, and the emissivity and absorption coefficient take their usual forms for the 21 cm line,
\begin{eqnarray}
    j_\nu & = & \frac{3}{16\pi} \frac{hc}{\lambda_0} n_\mathrm{H^0} A_{21} \phi_\nu \\
    \kappa_\nu & = & \frac{3}{32\pi} \frac{hc\lambda_0}{k_\mathrm{B} T_\mathrm{spin}} n_\mathrm{H^0} A_{21} \phi_\nu.
\end{eqnarray}
In these expressions $\lambda_0 = 21.12$ cm is the line centre wavelength, $A_{21} = 2.8843 \times 10^{-15}$ s$^{-1}$ is the Einstein $A$ coefficient, $T_\mathrm{spin}$ is the spin temperature, and
\begin{equation}
    \phi_\nu = \frac{1}{\sqrt{2\pi}} \frac{\lambda}{\sigma_V} \exp\left(-\frac{v^2}{2\sigma_V^2}\right)
\end{equation}
is the line shape function evaluated, with $\sigma_V$ the velocity dispersion, which we assume to be the isothermal sound speed of an ideal gas, and $v = c (1 - \nu \lambda_0/hc)$ the velocity corresponding to frequency $\nu$. In practice, we evaluate $I_\nu$ between the velocity range $-25$ $\kms$ to $25 \kms$ with a $0.1 \kms$ velocity resolution. Our procedure to calculate the spin temperature is the same as that of \citet{Kim2014}.

We calculate solutions for $I_0 = B_\nu(T_0)$ evaluated at both $T_0=3.77$ K and $T_0=10^5$ K, where $B_\nu(T)$ is the Planck function. The former corresponds to lines of sight where there is no background other than the cosmic microwave and radio backgrounds \citep{Kim2014}, while the latter represents a typical observation along a sightline toward a radio-bright quasar. We then convert the observed intensity as a function of frequency $I_\nu$ to an antenna temperature as a function of velocity via the usual conversion $T_A = c^2 I_\nu/2 k_B \nu^2$ evaluated at frequency $\nu = hc/\lambda_0 (1 - v/c)$.

\subsubsection{Derived quantities}
\label{ssec:derived_quantities}

From the computed observables, we can then compute a series of derived quantities. Our goal is to investigate how well one can predict DMs from these derived quantities. The most straightforward are emission measures, which we obtain by inverting the expressions above for the H$\alpha$ and free-free emissivity, assuming a constant temperature $T=10^4$ K and (for free-free) observing frequency $\nu = 1$ GHz. Specifically, we define the free-bound (H$\alpha$) emission measure by
\begin{equation}
    \mathrm{EM_{fb,obs}} = \frac{4 \pi I_{\mathrm{H\alpha}}}{\alpha_{\mathrm{H\alpha,eff}} h\nu_{\mathrm{H\alpha}}},
\end{equation}
where $\alpha_{\mathrm{H\alpha,eff}}$ is derived from \autoref{eq:alpha_eff} with $T_4 = 1$. Substituting the constants,
\begin{equation}
    \left( \frac{\mathrm{EM}_\mathrm{fb,obs}}{\pc \cm^{-6}} \right) = \left( \frac{4\pi\times2.75 \times 10^{-6}}{\cm^{-4} \pc \; \s \; \sr} \right)\frac{I_\mathrm{H\alpha} }{h\nu_\mathrm{H\alpha}},
\end{equation}
where we have divided by the value of $\pc$ in $\cm$ to obtain the conventional EM units of pc cm$^{-6}$. This expression is equivalent to equation 9 of \citet{Kaczmarek2017}. We follow the same process to derive EM from free-free emission as
\begin{equation}
    \mathrm{EM}_\mathrm{ff,obs} = \frac{4\pi I_\mathrm{ff}}{\mathcal{L}_{\nu,i,\mathrm{ff}}}.
\end{equation}
In the conventional EM units, we can rewrite this as
\begin{equation}
    \left( \frac{\mathrm{EM}_\mathrm{ff,obs}}{\pc \cm^{-6}} \right) = \left( \frac{4\pi \times 9.67 \times 10^{20}}{\erg^{-1} \cm^{-4} \; \sr \;  \pc \; \s \; \Hz} \right) I_\mathrm{ff}
\end{equation}
Note that the EMs derived from free-bound or free-free may differ from the true EM due to the assumption of constant temperature; the calculated emission uses the real gas temperature, whereas when inverting that emission to obtain EM, we are forced to assume a temperature.

We also derive a series of quantities from the 21 cm spectra. For the case without a background radio quasar (i.e., computed with $T_0 = 3.77$ K), we compute the inferred neutral hydrogen column density as
\begin{equation}
        N_\mathrm{\Hi,obs} = \left(\frac{10^{20}\;\mathrm{cm}^{-2}}{54.89\;\mathrm{K}\;\mathrm{km}\;\mathrm{s}^{-1}}\right)\int \left(T_A - T_0 \right) \, \mathrm{d}v
        ,
\end{equation}
which follows from assuming $\tau_\nu \ll 1$ in \autoref{eq:Inu}; this is the standard result for the relationship between antenna temperature and H~\textsc{i} column in the optically thin limit. As with the observationally-inferred EMs, this observationally-inferred \Hi~column density may differ from the true one because it is derived under the assumption of small optical depth, whereas the actual 21 cm spectrum is computed properly accounting for opacity effects, which are non-negligible along some sightlines (e.g., \citealt{Hu2023} show that sightlines towards high-density regions in the simulation show \Hi~self-absorption).

Finally, for the 21 cm spectrum including the effects of a background source that provides $T_0 = 10^5$ K, we follow the usual procedure \citep[e.g.,][]{Draine2011} for deriving the 21 cm optical depth and spin temperature from the difference between the ``on'' and ``off'' source spectra, i.e., those computed with $T_0 = T_{0,\mathrm{on}} = 10^5$ K and those computed with $T_0 = T_{0,\mathrm{off}}=3.77$ K. Specifically, we compute these as
\begin{eqnarray}
    \tau_{\nu}
        & = & \ln \left( \frac{T_{0, \mathrm{on}} - T_{0, \mathrm{off}}}{T_{A,\mathrm{on}} - T_{A,\mathrm{off}}} \right) \\
    T_\mathrm{spin} & = & \frac{T_{A,\mathrm{off}} T_{0,\mathrm{on}} - T_{A,\mathrm{on}} T_{0,\mathrm{off}}}{(T_{0,\mathrm{on}} - T_{0,\mathrm{off}}) - (T_{A,\mathrm{on} - T_{A,\mathrm{off}}})}.
\end{eqnarray}
where $T_{A,\mathrm{on}}$ and $T_{A,\mathrm{off}}$ are the antenna temperatures as a function of velocity computed using $T_0 = 10^5$ K and 3.77 K, respectively. This process yields an estimate of the optical depth and spin temperature in every velocity channel of the spectra. We then extract the maximum optical depth, $\tau_\mathrm{max}$, and the minimum spin temperature, $T_\mathrm{spin,min}$, to use in the remainder of our analysis. These two quantities are potentially interesting as diagnostics because they detect the presence of a cold atomic phase that might be missed in the 21 cm emission alone.

Thus for each sightline, we have the following collection of quantities: rotation measure RM, dispersion measure DM, true emission measure EM, true \Hi~column density, emission measures derived from H$\alpha$ and free-free emission, EM$_\mathrm{fb,obs}$ and EM$_\mathrm{ff,obs}$, \Hi~column density derived from 21 cm emission, $N_\mathrm{\Hi,obs}$, and maximum 21 cm optical depth and minimum spin temperature derived from emission plus absorption, $\tau_\mathrm{max}$ and $T_\mathrm{spin,min}$. Of these, six -- RM, EM$_\mathrm{fb,obs}$, EM$_\mathrm{ff,obs}$, $N_\mathrm{\Hi,obs}$, $\tau_\mathrm{max}$, and $T_\mathrm{spin,min}$ -- are generally observable, while the remaining three -- DM, EM, and $N_\mathrm{\Hi,true}$ will prove useful for the purpose of gaining physical insight, but are not generally accessible to observations.

\subsubsection{Sample generation and observational limits}
\label{sssec:samples}

Having introduced the procedure by which we compute observables and derived quantities for a specified LOS, our next step is to generate a large sample of LOS's. For this purpose we adopt a coordinate system in which the observer is located at $z=+\infty$, and we rotate each galaxy relative to the $xy$ plane to achieve a target inclination angle. For all results presented in the main text, we adopt a realistic inclination angle of $34.7^{\circ}$ for both LMC cases, and apply a similar inclination of $30^{\circ}$ to the MW case to maintain consistency; we explore the dependence of our results on inclination angle, and show that this dependence is mostly minor as long as the inclination angle is $<60^\circ$, in \aref{sec:robustness_tests}. We then define a square sampling region of $25 \ \mathrm{kpc} \times 25 \ \mathrm{kpc}$ for the LMC cases and $40 \ \mathrm{kpc} \times 40 \ \mathrm{kpc}$ for the MW case, within which we draw approximately 20000, 10000 and 3000 uniformly-distributed sightlines for the \LMCI\, \LMCW\, and \MWI\ cases, respectively. We discard sightlines, typically located in the galactic outskirts, where the gas particle density is too low to solve the radiative transfer equation along the entire ray continuously (see \aref{sec:AdaptiveBinning}). We show in \aref{sec:robustness_tests} that our samples are large enough to yield statistically-converged results.

Because we randomly select our sightlines, there is no guarantee that any of the observational tracers we compute for them will be in the range of plausible detectability. We must therefore impose some limits on our synthetic observable quantities to ensure that they are actually within the observable range. We take a fairly conservative approach to this culling process, generally keeping any sightline that would be detectable to the most sensitive studies in the literature.

For EMs -- both free-bound and free-free-based -- we consider a sightline undetectable if the EM is $<10^{-1} \pc \cm^{-6}$, consistent with typical observational sensitivities in these tracers \citep{Tufte1998, Haffner2001}. Then we adopt $\tau_\mathrm{max} = 10^{-3}$ as our maximum optical depth threshold, based on all-sky surveys like GASKAP, deep MeerKAT observations, and surveys using the Giant Metrewave Radio Telescope (GMRT) and the Westerbork Synthesis Radio Telescope (WSRT) \citep{Roy2013, Dickey2013, Nguyen2024}. Finally, we take $N_\mathrm{\Hi,obs} = 10^{17}$ cm$^{-2}$ as our minimum detectable neutral hydrogen column density threshold, selected from earlier deeper surveys using the Westerbork array, VLA, and WSRT \citep{Braun2004, Walter2008, Heald2011} and also newer surveys like MHONGOOSE using MeerKAT observations \citep{deblok2024}. 


For sightlines on which we flag one or more quantities as unobservable, we mask those values from the quantitative analysis that follows. Thus for example when we compute correlation coefficients between $\mathrm{EM}_\mathrm{fb,obs}$ and other quantities, we exclude sightlines with $\mathrm{EM}_\mathrm{fb,obs} < 0.1$ pc cm$^{-6}$ from these calculations. We only mask the unobservable quantity, not other quantities on the same sightline that are in the observable range.

\section{Results}
\label{sec:results}
In this section, we explore different mathematical models for predicting the average LOS magnetic fields $\Bz$ from RM measurements, where along each sightline we define
\begin{equation}
    \label{eq:Bzne}
    \Bz \equiv \frac{\int \ne B_s\, \mathrm{d}s}{\int \ne \,\mathrm{d}s}
\end{equation}
as the electron density-weighted mean LOS magnetic field. Since the dispersion measure is $\mathrm{DM} = \int\ne\, \mathrm{d}s$ and the rotation measure is $\mathrm{RM}=k_\mathrm{RM}\int \ne  B_s \, \mathrm{d}s$, by definition we have
\begin{equation}
    \Bz = k_\mathrm{RM}^{-1} \frac{\mathrm{RM}}{\mathrm{DM} } 
\end{equation}
and the problem of measuring the magnetic field from an RM reduces to that of obtaining the corresponding DM. Now we quantitatively investigate the predictive power of indirect observables (described in \autoref{ssec:derived_quantities}) in predicting the DMs.

\subsection{Correlation between DM and observables}
\label{ssec:DM-correlations}

\begin{figure*}
    \includegraphics[width=2\columnwidth]{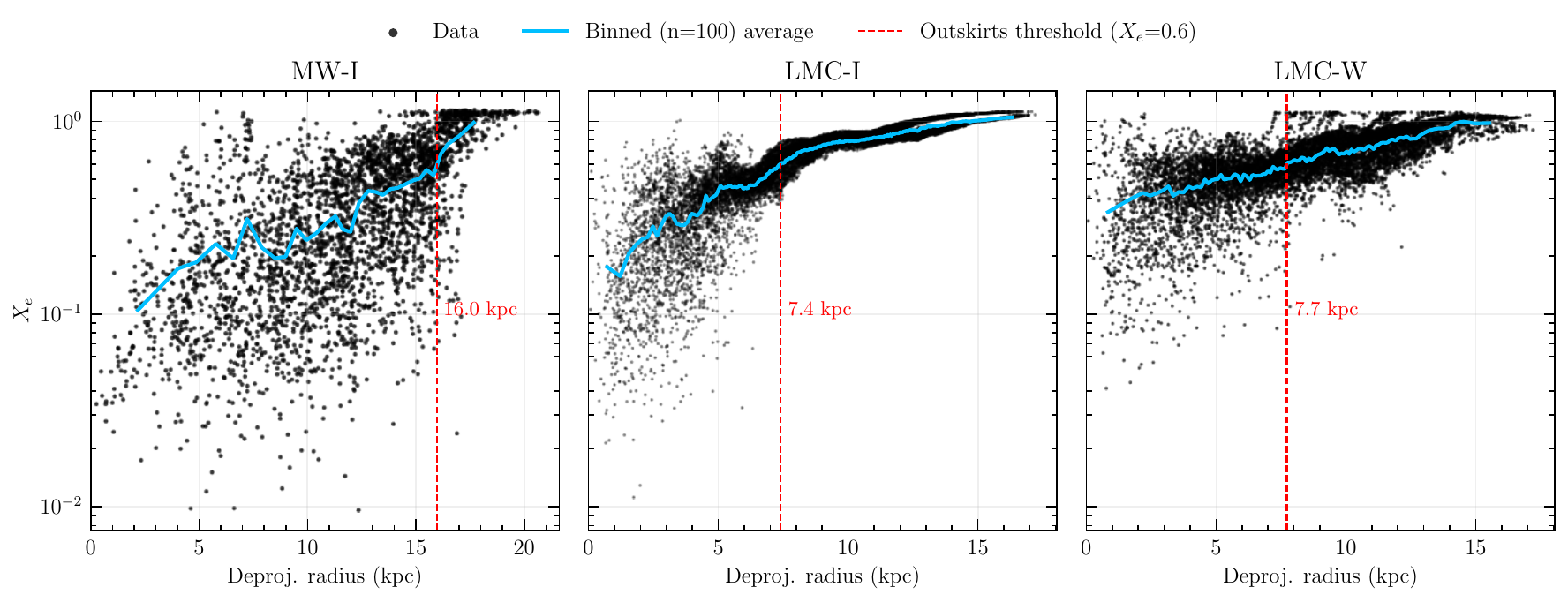} 
    \caption{Ionisation fractions $X_e$ of all sightlines through different galaxies -- \MWI, \LMCI, and \LMCW\ (columns) -- as a function of the deprojected radius. The blue line shows the average ionisation fraction in bins of 100 data points when sorted from minimum to maximum deprojected radii. The red vertical line indicates the deprojected radius at which the linearly interpolated average ionisation fraction is 0.6, our threshold to separate the central ISM from the outskirts.}
    \label{fig:Xe_radial_dependence}
\end{figure*}

Before we proceed to full mathematical models to predict DMs, we start with a simple investigation of how well our various observables correlate with DM. An important consideration for this purpose is that our sightlines probe a wide range of radii in our different simulations  ($\approx$0-21 kpc for \MWI\ and $\approx$0-17 kpc for \LMCI\ and \LMCW) and consequently probe a wide range of environments, from central discs where star formation and neutral (atomic and molecular) gas are abundant to outskirts dominated by ionised gas. For this reason, it is helpful to distinguish between the central disc ISM and the outskirts, and we therefore define the outskirts radius as the radius beyond which the mean electron fraction of all bins of sightlines (100 sightlines per bin) $\langle X_e\rangle > 0.6$,\footnote{We use $X_e = 0.6$ rather than 0.5 to separate inner and outer regions because $X_e$ spans the range 0 to $\approx 1.2$, with the latter case corresponding both H and He being fully ionised.} where for any individual sightline we define $X_e = \int n_e\,\mathrm{d}s / \int n_{\rm H}\,\mathrm{d}s$. We plot $X_e$ versus (deprojected) galactocentric radius for all sightlines in \autoref{fig:Xe_radial_dependence}. In this figure, cyan lines show the binned average, and the vertical red dashed lines indicate where this mean passes through 0.6. Interior to this radius, ionisation fractions exhibit substantial scatter, while beyond it most sightlines are primarily ionised, a distinction that will prove to be important below. Our analysis gives an outskirts radius of 16.0 kpc for \MWI\, and 7.4 and 7.7 kpc for \LMCI\ and \LMCW, respectively.

\begin{figure*}
    \includegraphics[width=2\columnwidth]{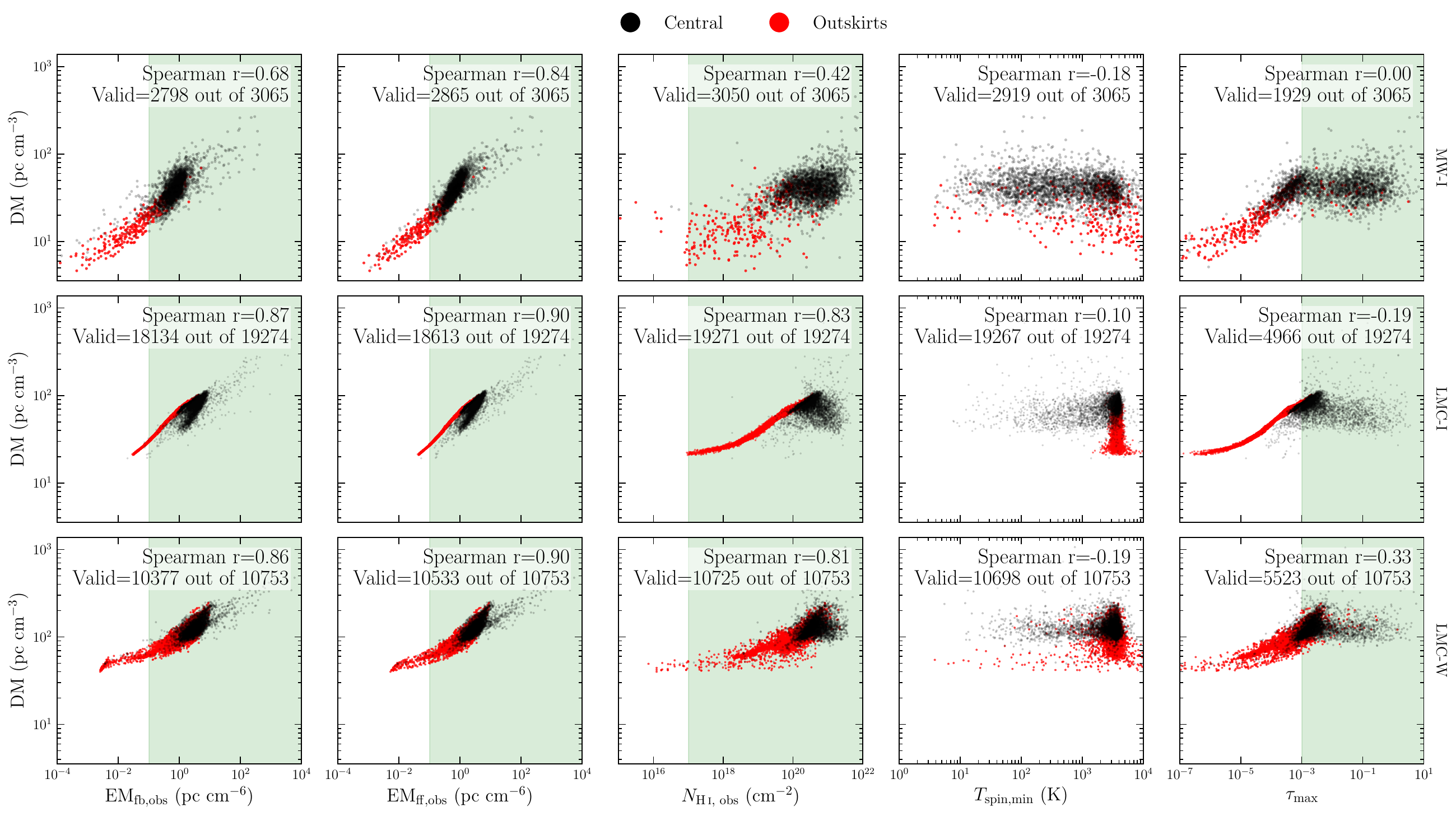} 
    \caption{Scatter plots of DM versus EM$_\mathrm{fb,obs}$, EM$_\mathrm{ff,obs}$, $N_\mathrm{\Hi,obs}$, $T_{\rm spin,min}$, and $\tau_\mathrm{max}$ (columns, left to right) for the galaxies \MWI, \LMCI, and \LMCW\ (rows, top to bottom). Black points show sightlines through the central ISM, and red points show sightlines through the outskirts. Green vertical bands mark the range of parameter values that we consider detectable (see \autoref{sssec:samples}). The panel text lists the total Spearman rank $r$ of the correlation and both the total number of simulated sightlines and the number that are detectable (``valid''), with only the latter included in the calculation of $r$. \autoref{tab:spearman_correlations} lists the separate Spearman ranks for the central and outskirts regions. \\}
    \label{fig:DM_correlations}
\end{figure*}

With the distinction between inner and outer galaxy regions established, we now show in \autoref{fig:DM_correlations} scatter plots of DM versus our five observable quantities, separated by whether the sightlines are in central regions (black) or the outskirts (red). We report the Spearman ranks of the correlations (combining both regions) in each panel, and we provide Spearman ranks for the inner and outer disk regions separately in \autoref{tab:spearman_correlations}. From the plots and the Spearman ranks, it is clear that across all three simulations and in both inner and outer regions, EMs exhibit the highest correlation with DMs and $\Tspin$ the lowest. 

The correlations of DM with $N_\mathrm{\Hi,obs}$ are moderate for the galaxies taken as a whole, but from the differences between the black and red points in the figure, and the Spearman ranks broken out by inner versus outer galaxy in \autoref{tab:spearman_correlations}, it is clear that this correlation is significantly stronger in the outskirts compared to the central regions. The same is true of $\tau_\mathrm{max}$ in \LMCI~ and \LMCW, but not in \MWI. This trend has a simple physical interpretation: in the central disc, the neutral and ionised phases are largely separated in space, and neutral gas properties such as $N_{\rm HI,obs}$ and $\tau_\mathrm{max}$ provide little information about the ionised gas, which is the phase traced by DM. By contrast, in the outskirts -- where neutral gas is sparse -- the neutral gas exists primarily as a sub-dominant tracer that is spatially well-mixed and dynamically coupled to the dominant ionised phase; tracers of the neutral gas therefore become informative about the ionised gas. The only reason that $\tau_\mathrm{max}$ performs less well than $N_{\rm HI,obs}$ in the outskirts for \MWI~is that the range of $\tau_\mathrm{max}$ over which there is a good correlation with DM is too small to be detectable due to the overall larger optical depth of \Hi~in the higher-density \MWI~simulation. Because our simulations (with the exception of \LMCW) lack a CGM, they are not primarily designed to probe galactic outskirts. The \MWI\ simulation forms a highly compact disc devoid of gas in the extended outskirt regions, which is not the case with both LMC simulations. Thus, we rely more heavily on the LMC simulations to analyse differences between the central ISM and the outskirts, and expect that the qualitative results apply to all our simulated cases.


\begin{table*}
\centering
\caption{Spearman rank correlation coefficients for DM correlations. Values show central (outskirts) regions for observationally valid sightlines.}
\label{tab:spearman_correlations}
\begin{tabular}{lccccc}
\hline
Galaxy & \shortstack{DM \\ vs. \\ EM$_{\mathrm{fb,obs}}$} & \shortstack{DM \\ vs. \\ EM$_{\mathrm{ff,obs}}$} & \shortstack{DM \\ vs. \\ $N_{\mathrm{HI,obs}}$} & \shortstack{DM \\ vs. \\ $T_{\mathrm{spin,min}}$} & \shortstack{DM \\ VS \\ $\tau_{\mathrm{max}}$} \\
\hline
\MWI & 0.66 (0.87) & 0.83 (0.93) & 0.28 (0.70) & $-0.13$ ($-0.48$) & 0.01 ($-0.26$) \\
\LMCI & 0.68 (0.99) & 0.77 (0.99) & 0.08 (0.98) & 0.31 (0.21) & $-0.19$ (0.85) \\
\LMCW & 0.71 (0.84) & 0.78 (0.88) & 0.39 (0.82) & $-0.03$ ($-0.04$) & 0.19 (0.50) \\
\hline
\end{tabular}
\end{table*}


\subsection{Models for predicting magnetic fields}
\label{ssec:models}

Based on our analysis of the correlations between DMs and observables, it seems that EM -- measured either from free-free or bound-free -- and $N_\mathrm{\Hi,obs}$ are the most promising probes for the indirect measurement of DMs, and thus of magnetic fields. To define a model that can predict DM from these probes, we consider double power-law functional fits of the form
\begin{equation}
    \label{eq: DM_predict}
    \mathrm{DM}_\mathrm{pred} = K \, \mathrm{EM}^\alpha \, N_\mathrm{\Hi,obs}^\beta,
\end{equation}
where $\alpha$, $\beta$, and $K$ are parameters to be fit, and $K$ has units of $(\mathrm{length})^{5 \alpha + 2 \beta - 2}$. Equivalently, we can write the LOS-averaged magnetic field predicted by this model as
\begin{equation}
    \Bpredicted = k_\mathrm{RM}^{-1} K^{-1} \mathrm{RM} \, \mathrm{EM}^{-\alpha} N_\mathrm{\Hi,obs}^{-\beta}.
    \label{eq:Bpred}
\end{equation}
Taking the logarithm (base 10 throughout the paper)\footnote{For this step, we take the absolute value of $\Bpredicted$ and RM. There is no information loss in doing so because the sign of LOS magnetic field is determined solely by the sign of RM, as all the other quantities appearing in \autoref{eq:Bpred} are positive-definite.}, converting to conventional units for the EMs, and normalising $\NH$ with a factor of $10^{20} \cm^{-2}$, we have
\begin{eqnarray}
\lefteqn{\log\left(\frac{|\Bpredicted|}{\mathrm{\mu G}}\right)
= } \notag \\
& \quad &
  -\log\left( \frac{k_{\mathrm{RM}}}{\mathrm{rad\, m^{-2}} \,(\mathrm{pc}\,\mathrm{cm}^{-3})^{-1}} \right) - \log\left( \frac{K}{\pc^{1-\alpha}\mathrm{cm}^{6\alpha + 2\beta - 3}} \right) + {}\notag \\
&&
  \log\left( \frac{\mathrm{|RM|}}{\mathrm{rad\,m^{-2}}} \right) - \alpha \log\left( \frac{\mathrm{EM}}{\pc\ \mathrm{cm^{-6}}} \right) - \beta \log\left( \frac{N_\mathrm{\Hi,obs}}{10 ^{20}\mathrm{cm^{-2}}} \right),
\label{eq:logB_predict_conventional}
\end{eqnarray}
For our subsequent results we will use the conventional units for better readability, unless stated otherwise.

The remaining task then is to measure the free parameters $\alpha$, $\beta$, and $\log K$ of the model. For this purpose we use \textsc{scipy.optimize.curve\_fit}\footnote{\url{https://docs.scipy.org/doc/scipy/reference/generated/scipy.optimize.curve_fit.html}} to find the best non-linear least squares fit for the model given by \autoref{eq:logB_predict_conventional}, using as input data the true values of $\Bz$ and the observables RM, EM, and $N_\mathrm{\Hi, obs}$. We carry out several variations of this fit. First, for EM we analyse both EM$_\mathrm{fb}$ and EM$_\mathrm{ff}$ (i.e., EMs derived from both optical and radio data) in order to check whether one or the other performs better. Second, we also carry out fits using the true values of the EM and \Hi~column density rather than their observable proxies, which enables us to distinguish between errors due to a failure of the underlying correlation (i.e., EM and \Hi~column do not predict DM that well even if they are perfectly measured) and errors due to our inability to measure the observables (i.e., our observational estimates of EM do not reflect the true value of EM, and this causes errors when we use them to predict DM). Third, we consider both models in which $\alpha$ and $\beta$ are each left free, and those with either $\alpha$, $\beta$, or both fixed to zero. The case where $\alpha$ and $\beta$ are both free corresponds to a situation where an observer has access to both measurements of EM and $N_\mathrm{\Hi,obs}$, while the cases where either $\alpha$, $\beta$, or both are fixed to zero correspond to situations where one or both of the observables are unavailable and thus the DM must be deduced from reduced information. In particular, the case $\alpha=\beta=0$ corresponds to the simplest possible approach of assuming a fixed DM for all sightlines, as has sometimes been done in the earlier observational literature \citep{Gaensler2005}.

\begin{figure*}
    \includegraphics[width=2\columnwidth]{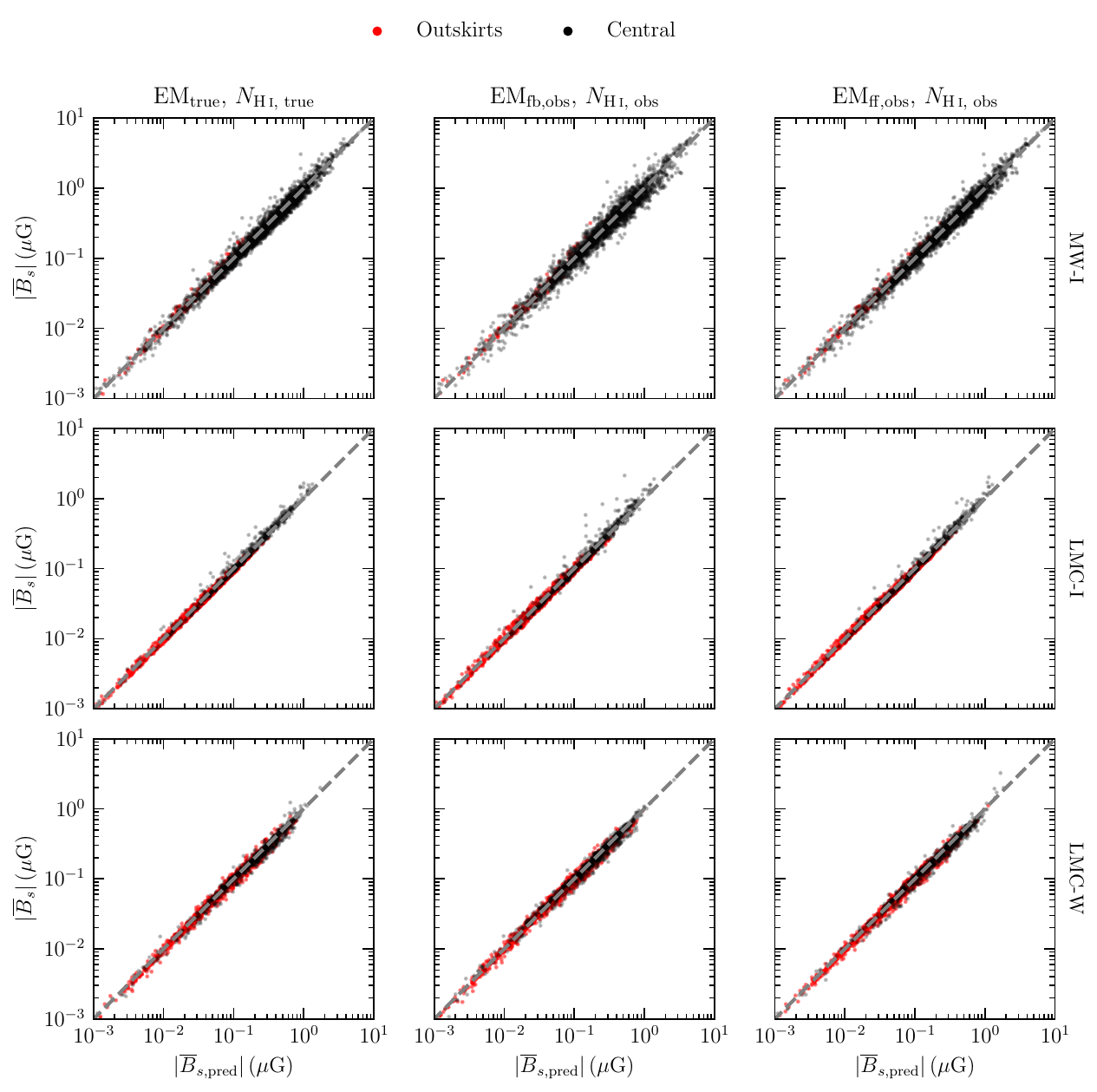} 
    \caption{True LOS-averaged magnetic field strength $|\Bz|$ versus predicted strength $|\Bpredicted|$ using best-fitting values of the fit parameters $\alpha$, $\beta$, and $K$, for our three different simulations (\MWI, \LMCI, \LMCW -- top to bottom rows) and for three sets of inputs (true EM and \Hi~column, left; optical-derived EM$_\mathrm{fb}$ and observationally-inferred \Hi~column $N_\mathrm{\Hi,obs}$, middle; radio-derived EM$_\mathrm{ff}$ and observationally-inferred \Hi~column $N_\mathrm{\Hi,obs}$; right). Grey-black points show inner-galaxy sightlines, red points show outer-galaxy sightlnes, and grey dashed lines show the one-to-one relation.}
    \label{fig:Btrue_Bmeasured}
\end{figure*}

\begin{table}
\centering
\footnotesize
\setlength{\tabcolsep}{3pt}
\caption{Best-fit parameters for the unconstrained (free) model. We report fits for every simulation (first column), separately for the central (C) and outskirts (O) regions (second column), and each combination of observables -- true \Hi~column and EM, observationally-derived \Hi-column and EM from free-free, and observationally-derived \Hi-column and EM from free-bound (column 3). For each model combination, the table reports the best-fitting values of the parameters $\alpha$, $\beta$, and $K$ (columns 4-6) and the scatter $\sigma_{\chi}$ (column 7). Note that $K$ is in conventional units of pc$^{1-\alpha}$ cm$^{6\alpha+2\beta-3}$ (see \autoref{eq:logB_predict_conventional} for more details).}
\label{tab:missing_observations_free}
\begin{tabular}{@{}c c c@{\hspace{0.2in}} c c c c@{}}
\hline
\multicolumn{3}{c}{Model} &
\multicolumn{4}{c}{Fit result} \\
Sim & Reg & Obs & $\alpha$ & $\beta$ & $\log K$ & $\sigma_{\chi}\,(\mathrm{dex})$ \\
\hline
\multirow{6}{*}{\MWI} & \multirow{3}{*}{C} & $N_\mathrm{\Hi, true}$, EM & 0.34 & -0.00 & 1.64 & 0.078 \\
 &  & $N_\mathrm{\Hi, obs}$, $\mathrm{EM_{ff}}$ & 0.33 & 0.00 & 1.66 & 0.085 \\
 &  & $N_\mathrm{\Hi, obs}$, $\mathrm{EM_{fb}}$ & 0.26 & 0.00 & 1.65 & 0.105 \\
\cline{2-7}
 & \multirow{3}{*}{O} & $N_\mathrm{\Hi, true}$, EM & 0.42 & 0.01 & 1.64 & 0.052 \\
 &  & $N_\mathrm{\Hi, obs}$, $\mathrm{EM_{ff}}$ & 0.38 & 0.01 & 1.64 & 0.052 \\
 &  & $N_\mathrm{\Hi, obs}$, $\mathrm{EM_{fb}}$ & 0.32 & 0.00 & 1.63 & 0.061 \\
\hline
\multirow{6}{*}{\LMCI} & \multirow{3}{*}{C} & $N_\mathrm{\Hi, true}$, EM & 0.34 & -0.09 & 1.77 & 0.058 \\
 &  & $N_\mathrm{\Hi, obs}$, $\mathrm{EM_{ff}}$ & 0.33 & -0.10 & 1.78 & 0.057 \\
 &  & $N_\mathrm{\Hi, obs}$, $\mathrm{EM_{fb}}$ & 0.31 & -0.14 & 1.81 & 0.061 \\
\cline{2-7}
 & \multirow{3}{*}{O} & $N_\mathrm{\Hi, true}$, EM & 0.41 & -0.02 & 1.79 & 0.011 \\
 &  & $N_\mathrm{\Hi, obs}$, $\mathrm{EM_{ff}}$ & 0.39 & -0.01 & 1.81 & 0.012 \\
 &  & $N_\mathrm{\Hi, obs}$, $\mathrm{EM_{fb}}$ & 0.33 & 0.01 & 1.84 & 0.014 \\
\hline
\multirow{6}{*}{\LMCW} & \multirow{3}{*}{C} & $N_\mathrm{\Hi, true}$, EM & 0.22 & 0.03 & 1.97 & 0.057 \\
 &  & $N_\mathrm{\Hi, obs}$, $\mathrm{EM_{ff}}$ & 0.21 & 0.02 & 1.99 & 0.058 \\
 &  & $N_\mathrm{\Hi, obs}$, $\mathrm{EM_{fb}}$ & 0.20 & 0.00 & 2.00 & 0.063 \\
\cline{2-7}
 & \multirow{3}{*}{O} & $N_\mathrm{\Hi, true}$, EM & 0.36 & -0.07 & 1.96 & 0.047 \\
 &  & $N_\mathrm{\Hi, obs}$, $\mathrm{EM_{ff}}$ & 0.30 & -0.05 & 1.98 & 0.048 \\
 &  & $N_\mathrm{\Hi, obs}$, $\mathrm{EM_{fb}}$ & 0.21 & -0.00 & 1.99 & 0.052 \\
\hline
\end{tabular}
\end{table}

\begin{table}
\centering
\footnotesize
\setlength{\tabcolsep}{3pt}
\caption{Same as \autoref{tab:missing_observations_free}, but for the constrained model with $\alpha = 0$. Note that, since this model does not depend on EM, we do not compare different ways of estimating EM as we do for the free model in \autoref{tab:missing_observations_free}.}
\label{tab:missing_observations_alpha0}
\begin{tabular}{@{}c c c@{\hspace{0.2in}} c c c c@{}}
\hline
\multicolumn{3}{c}{Model} &
\multicolumn{4}{c}{Fit result} \\
Sim & Reg & Obs & $\alpha$ & $\beta$ & $\log K$ & $\sigma_{\chi}\,(\mathrm{dex})$ \\
\hline
\multirow{4}{*}{\MWI} & \multirow{2}{*}{C} & $N_\mathrm{\Hi, true}$ & 0 & 0.10 & 1.56 & 0.145 \\
 &  & $N_\mathrm{\Hi, obs}$ & 0 & 0.09 & 1.57 & 0.146 \\
\cline{2-7}
 & \multirow{2}{*}{O} & $N_\mathrm{\Hi, true}$ & 0 & 0.17 & 1.44 & 0.188 \\
 &  & $N_\mathrm{\Hi, obs}$ & 0 & 0.17 & 1.45 & 0.189 \\
\hline
\multirow{4}{*}{\LMCI} & \multirow{2}{*}{C} & $N_\mathrm{\Hi, true}$ & 0 & 0.01 & 1.87 & 0.109 \\
 &  & $N_\mathrm{\Hi, obs}$ & 0 & 0.02 & 1.86 & 0.109 \\
\cline{2-7}
 & \multirow{2}{*}{O} & $N_\mathrm{\Hi, true}$ & 0 & 0.23 & 1.85 & 0.026 \\
 &  & $N_\mathrm{\Hi, obs}$ & 0 & 0.23 & 1.86 & 0.026 \\
\hline
\multirow{4}{*}{\LMCW} & \multirow{2}{*}{C} & $N_\mathrm{\Hi, true}$ & 0 & 0.12 & 2.03 & 0.090 \\
 &  & $N_\mathrm{\Hi, obs}$ & 0 & 0.12 & 2.03 & 0.090 \\
\cline{2-7}
 & \multirow{2}{*}{O} & $N_\mathrm{\Hi, true}$ & 0 & 0.16 & 1.98 & 0.064 \\
 &  & $N_\mathrm{\Hi, obs}$ & 0 & 0.16 & 1.99 & 0.064 \\
\hline
\end{tabular}
\end{table}

\begin{table}
\centering
\footnotesize
\setlength{\tabcolsep}{3pt}
\caption{Same as \autoref{tab:missing_observations_free}, but for the constrained model with $\beta = 0$. Note that, since this model does not depend on \Hi~column, we do not compare different ways of estimating $\NH$ as we do for the free model in \autoref{tab:missing_observations_free}.}
\label{tab:missing_observations_beta0}
\begin{tabular}{@{}c c c@{\hspace{0.2in}} c c c c@{}}
\hline
\multicolumn{3}{c}{Model} &
\multicolumn{4}{c}{Fit result} \\
Sim & Reg & Obs & $\alpha$ & $\beta$ & $\log K$ & $\sigma_{\chi}\,(\mathrm{dex})$ \\
\hline
\multirow{6}{*}{\MWI} & \multirow{3}{*}{C} & EM & 0.34 & 0 & 1.64 & 0.078 \\
 &  & $\mathrm{EM_{ff}}$ & 0.33 & 0 & 1.66 & 0.085 \\
 &  & $\mathrm{EM_{fb}}$ & 0.27 & 0 & 1.65 & 0.105 \\
\cline{2-7}
 & \multirow{3}{*}{O} & EM & 0.44 & 0 & 1.65 & 0.053 \\
 &  & $\mathrm{EM_{ff}}$ & 0.39 & 0 & 1.65 & 0.052 \\
 &  & $\mathrm{EM_{fb}}$ & 0.32 & 0 & 1.63 & 0.061 \\
\hline
\multirow{6}{*}{\LMCI} & \multirow{3}{*}{C} & EM & 0.30 & 0 & 1.73 & 0.064 \\
 &  & $\mathrm{EM_{ff}}$ & 0.28 & 0 & 1.74 & 0.065 \\
 &  & $\mathrm{EM_{fb}}$ & 0.24 & 0 & 1.76 & 0.073 \\
\cline{2-7}
 & \multirow{3}{*}{O} & EM & 0.39 & 0 & 1.80 & 0.011 \\
 &  & $\mathrm{EM_{ff}}$ & 0.37 & 0 & 1.81 & 0.012 \\
 &  & $\mathrm{EM_{fb}}$ & 0.34 & 0 & 1.84 & 0.014 \\
\hline
\multirow{6}{*}{\LMCW} & \multirow{3}{*}{C} & EM & 0.22 & 0 & 1.98 & 0.057 \\
 &  & $\mathrm{EM_{ff}}$ & 0.22 & 0 & 1.99 & 0.058 \\
 &  & $\mathrm{EM_{fb}}$ & 0.20 & 0 & 2.00 & 0.063 \\
\cline{2-7}
 & \multirow{3}{*}{O} & EM & 0.26 & 0 & 1.97 & 0.048 \\
 &  & $\mathrm{EM_{ff}}$ & 0.24 & 0 & 1.98 & 0.049 \\
 &  & $\mathrm{EM_{fb}}$ & 0.21 & 0 & 1.99 & 0.052 \\
\hline
\end{tabular}
\end{table}

\begin{table}
\centering
\footnotesize
\setlength{\tabcolsep}{3pt}
\caption{Same as \autoref{tab:missing_observations_free}, but for the constrained model with $\alpha = \beta = 0$. Note that this model does not make use of any observables.}
\label{tab:missing_observations_alphabeta0}
\begin{tabular}{@{}c c c@{\hspace{0.2in}} c c c c@{}}
\hline
\multicolumn{3}{c}{Model} &
\multicolumn{4}{c}{Fit result} \\
Sim & Reg & Obs & $\alpha$ & $\beta$ & $\log K$ & $\sigma_{\chi}\,(\mathrm{dex})$ \\
\hline
\multirow{2}{*}{\MWI} & C & --- & 0 & 0 & 1.61 & 0.156 \\
\cline{2-7}
 & O & --- & 0 & 0 & 1.29 & 0.256 \\
\hline
\multirow{2}{*}{\LMCI} & C & --- & 0 & 0 & 1.87 & 0.109 \\
\cline{2-7}
 & O & --- & 0 & 0 & 1.73 & 0.133 \\
\hline
\multirow{2}{*}{\LMCW} & C & --- & 0 & 0 & 2.10 & 0.099 \\
\cline{2-7}
 & O & --- & 0 & 0 & 1.98 & 0.110 \\
\hline
\end{tabular}
\end{table}

We report the best-fit parameters for our model with all parameters free in \autoref{tab:missing_observations_free}. The corresponding results with $\alpha$ and $\beta$ fixed to zero appear in \autoref{tab:missing_observations_alpha0} and \autoref{tab:missing_observations_beta0}, respectively, and the results where $\alpha$ and $\beta$ are both fixed to zero and only $K$ is allowed to vary in \autoref{tab:missing_observations_alphabeta0}. For each model variation, in addition to reporting the best-fit parameters, we also quantify how well that model performs by computing the RMS error
\begin{equation}
    \echi = \left[ \frac{1}{N_{\rm LOS}} \sum_{N_{\rm LOS}} (\log\Bz - \log\Bpredicted)^2 \right]^{1/2},
\end{equation}
where $N_\mathrm{LOS}$ is the number of sightlines fit. To visualise the error measured by $\echi$, in \autoref{fig:Btrue_Bmeasured} we show a comparison between the absolute magnitudes of the true $\Bz$ and the model-predicted value $\Bpredicted$ for the best-fitting model parameters for a subset of our models. The case shown is where $\alpha$ and $\beta$ are both left free (\autoref{tab:missing_observations_free}), and for this case, we show all three simulations (rows). The three columns compare measurements using the true values of EM and \Hi~column (left column) to those derived using EMs inferred from optical (middle; EM$_\mathrm{fb}$) and radio (right; EM$_\mathrm{ff}$) data and \Hi~column derived assuming the optically thin limit ($N_\mathrm{\Hi,obs}$). In these figures, $\echi$ simply quantifies the degree of scatter from the one-to-one relation. 

Examining \autoref{tab:missing_observations_free} - \autoref{tab:missing_observations_alphabeta0}, we can make a few general remarks. First, for the free-parameter model corresponding to the case when both EM and \Hi~measurements are available, we find that our best-fitting models give $\echi \lesssim 0.1$ dex (26\% fractional error) for all simulations and regardless of whether we use EMs derived from radio or optical data. The free-free-based EMs yield somewhat lower RMS errors than the bound-free-based ones, likely due to the weaker dependence of EM on electron temperature in free-free emission compared to free-bound emission (see \autoref{eq:alpha_eff}), but the difference in error is minimal, $\lesssim 0.02$ dex, as is the difference in error between using true and observationally-inferred EMs and \Hi~columns. This implies that observational errors in measuring EM or $\NH$ are not major contributors to the total error budget, and that, if an observer can derive EM and $\NH$ reasonably accurately, they can expect to be able to obtain magnetic field measurements to accuracies of a few tens of per cent regardless of whether the EMs come from radio or optical data.

Second, we note that, when $\alpha$ and $\beta$ are both left free, the best-fit value of $\beta$ is always very close to zero. Consistent with this, $\echi$ values for the free and $\beta = 0$ models are very similar. We conclude from this that EM is a much better predictor of electron density columns than $\NH$, and that if EM observations are available $\NH$ can essentially be ignored. By contrast, the $\alpha = 0$ (no EM) model performs worse than the $\beta = 0$ (no $\NH$) model in all regions, though the difference is smaller in the outskirts than in central regions. This is consistent with our observation in \autoref{ssec:DM-correlations} that $\NH$ can be a reasonable predictor of DM in outer parts of galaxies where the gas is mainly ionised.

Third, there are noticeable differences between both the best-fit values and the levels of error between our three simulations and between central and outskirts regions. Best-fit values of $\alpha$ are higher for isolated cases -- 0.34 for \MWI~and 0.34 for 
\LMCI{} -- than for the non-isolated case, \LMCW~(0.22). Similarly, $\alpha$ is always higher in the outskirts than in the central regions, while $\echi$ is consistently lower in the outskirts, indicating that magnetic field predictions are more reliable in the outer regions. For the $\alpha = \beta = 0$ model, $\log K$ (which in this case is simply the fixed value of DM that one should adopt) is lowest for \MWI, highest for \LMCW, and intermediate for \LMCI. This suggests that \MWI~has the most compact disk and \LMCW~the most diffuse, with \LMCI~lying in between. 

Fourth and finally, across all models, the highest RMS error $\echi \lesssim 0.25$ dex (fractional error $\approx 78\%$). This implies that, even with poor access to additional observables to calibrate DM, with appropriate free parameter choices magnetic fields can be predicted with reasonable accuracy. 


\section{Discussion}
\label{sec:discussion}
In this section, we explain the physics driving the relations seen in \autoref{sec:results}, discuss whether an observer can reliably estimate mass-weighted magnetic fields (as opposed to the electron density-weighted means we have investigated thus far), compare our models to those described in the literature, and finally present the recipe for cooking the best magnetic field values out of RM data.

\subsection{Physical interpretation of the free parameters}
\label{sec:physical_interpretation}
Here we explain the physical significance of the free parameters -- $\alpha$, $\beta$, and $K$ -- used to predict the DM (see \autoref{eq: DM_predict}). To begin this discussion, we consider an idealised case in which the electron density is non-zero only over a finite path length $L_p$ along the line of sight and thus we can write the dispersion and emission measures as $\mathrm{DM} = \langle n_e \rangle L_p$ and $\mathrm{EM} = \langle n_e^2 \rangle L_p$, where the angle brackets indicate averages over $L_p$. For this case, the relationship between DM and EM can then be expressed as
\begin{equation}
    \label{eq: DM_predict_fc}
    \mathrm{DM} = f_c^{-1/2} L_p^{1/2} \mathrm{EM}^{1/2}
\end{equation}
where we have introduced the clumping factor $f_c = \langle \ne^2\rangle / \langle \ne\rangle^2$ \citep{Gnedin1997} to quantify the amount of small-scale structure along a LOS. Comparing this expression to our empirical prediction model, \autoref{eq: DM_predict}, it is clear that if $L_p$ and $f_c$ were both constant across the face of a galaxy, then we would simply have $\alpha = 1/2$ and $\beta = 0$ -- and as we discuss in \autoref{ssec:comparison} this is indeed one of the empirical models in the literature that has been adopted in the past. In practice we expect the path length through the ionised gas $L_p$ to be roughly constant across different sightlines through an intervening extragalactic source, and thus we can intuitively view deviations of $\alpha$, $\beta$, and $K$ from these values as describing how the clumping factor $f_c$ varies from one sightline to another, and how well we can predict these variations from measurements of EM and $\NH$.

\begin{figure*}
    \includegraphics[width=2\columnwidth]{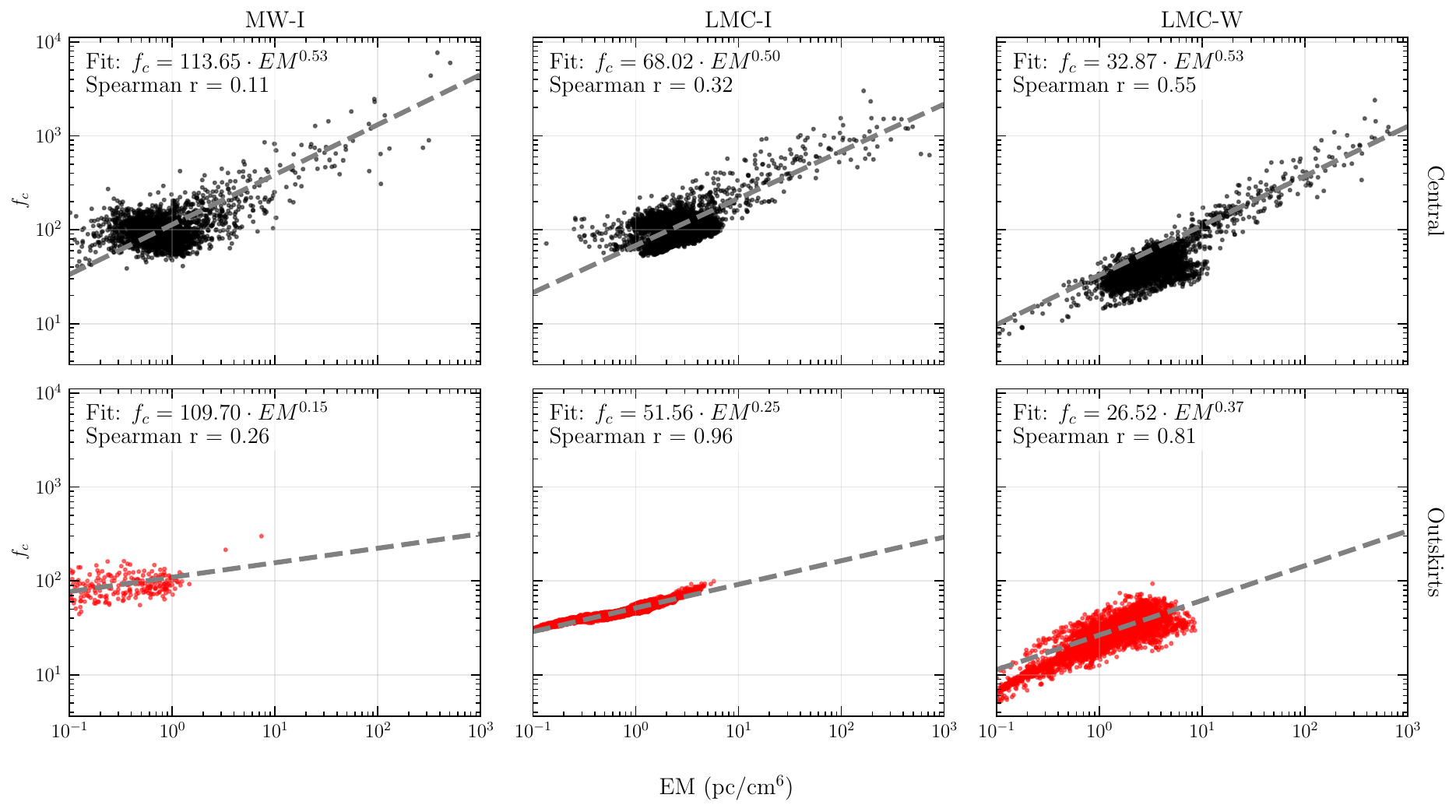} 
    \caption{Relationship between emission measure EM and clumping factor $f_c$ for all sightlines. The three columns show each of our three simulations, while the top and bottom rows show the central and outskirts regions, respectively. Annotations in each panel indicate the best linear fit and the Spearman rank correlation of the data shown, and the grey dashed lines show the best linear fit.}
    \label{fig:EM_fc_relations_central_outskirts}
\end{figure*}

\subsubsection{Explaining $\alpha$}
\label{sec:explaining_alpha}
The exponent $\alpha$ on the EM determines how strongly DM variations depend on EM variations. When $f_c$ is constant or uncorrelated with EM across multiple sightlines, $\alpha$ is close to 0.5. Conversely, when $f_c \propto \mathrm{EM}$, $\alpha$ approaches 0. The former situation prevails in a galaxy where EM variations arise primarily from changes in the total number of electrons (i.e., electron column density or DM) along each sightline, while the latter occurs if EM variations from one sightline to another are mainly driven by changes in the clumpiness of the electron distribution, with little change in total electron column density. Finally, for the more general case where $f_c$ is correlated with EM but not directly proportional to it, if we have a generic scaling $f_c \propto \mathrm{EM}^\gamma$, then we expect to obtain $\alpha = (1 - \gamma)/2$ (cf. \autoref{eq: DM_predict} and \autoref{eq: DM_predict_fc}). To check this, we compute $f_c$ along the same sightlines for which we compute EM, and we show the relationship between the two quantities in \autoref{fig:EM_fc_relations_central_outskirts}. We then compute the Spearman rank correlation and find the best-fit slope for the data shown in each panel of the figure. We see that $f_c$ and EM are highly correlated (Spearman ranks $>0.5$) for the \LMCW-central, \LMCI-outskirts, and \LMCW-outskirts cases, and for the best-fit values of the slope $\gamma$ obtained in these cases, the corresponding predicted slopes $\alpha$ are 0.24, 0.38, and 0.32, respectively. These are consistent with the values derived from curve fitting in \autoref{tab:missing_observations_free}, where $\alpha$ equals 0.22, 0.41, and 0.36, respectively. For the remaining cases, the Spearman rank correlation between $f_c$ and EM is low ($\lesssim 0.3$), and we therefore expect values of $\alpha$ relatively close to $0.5$ -- which is indeed what we do find: 0.34 for \MWI-central, 0.34 for \LMCI-central, and 0.42 for \MWI-outskirts. These trends therefore support our qualitative understanding of $\alpha$: when EM variations arise from electron redistribution (variations in clumpiness), $\alpha$ has a smaller value; when variations are driven by total electron amount, $\alpha$ increases.


\subsubsection{Explaining $\beta$}

The exponent $\beta$ describes how well one can predict EMs from $\NH$, and we have seen that $\beta\approx 0$ whenever EM data is available (i.e., when we do not fix $\alpha = 0$), consistent with our simple fixed-$L_p$ model. However, we have also seen that $\NH$ can be a useful predictor of DM in galaxy outskirts and when EMs are not available. We can understand this as follows: in the central ISM, there is a large amount of neutral gas along with some ionised gas produced by photoionisation from massive stars (\Hii\ regions), supernova‑heated thermally ionised gas, and various other dynamically uncorrelated processes. As a result, the neutral and ionised phases, and therefore their observational signatures, are spatially uncorrelated with each other, and $\NH$ is not useful as a predictor of DM ($\beta \approx 0$). In the outskirts, by contrast, the majority of the gas is ionised, and in this region \autoref{fig:Xe_radial_dependence} provides us with a useful insight into why $\beta \neq 0$. If the average ionisation fraction along a sightline is $X_e$, then we immediately have $\mathrm{DM} = \NH X_e/(1-X_e)$, so the predictive power of $\NH$ for DM therefore depends on how constant $X_e$ is from sightline to sightline. Clearly, from \autoref{fig:Xe_radial_dependence}, $X_e$ is much more constant across various sightlines in the outskirts, while it is highly variable in the central ISM. Thus, if one tries to estimate DM from $\NH$ in central regions, the variability in $X_e$ induces large uncertainties in DM. In the outskirts, where $X_e$ is roughly constant across several sightlines, variability in $X_e$ is much less important, and a model with moderate $\beta \approx 0.2$ outperforms one with $\beta = 0$.


\subsubsection{Explaining $K$}


The last free parameter, $K$, appears as the coefficient in \autoref{eq: DM_predict}, and combining this equation with our expression for DM in terms the clumping factor (\autoref{eq: DM_predict_fc}) gives
\begin{equation}
    \label{eq:logK}
    K = f_c^{-\alpha} L_p^{1-\alpha} \langle \ne\rangle^{1-2\alpha}.
\end{equation}
Thus, for a particular choice of $\alpha$, $K$ is a combination of the average (across multiple sightlines) clumping factors, electron densities, and path lengths, making it challenging to interpret. We can, however, gain some insight by considering the limiting cases $\alpha = 0$ and $\alpha = 1/2$. From \autoref{eq:logK}, when $\alpha = 0$, $K$ reduces to $\langle n_e \rangle L_p$ and thus represents the average DM\footnote{The value of $K$ varies marginally as a function of $\alpha$ and $\beta$ due to our choice of units and normalisations in \autoref{eq:logB_predict_conventional}.}. Thus the best-fitting values for our $\alpha = \beta = 0$ models (\autoref{tab:missing_observations_alphabeta0}) for our three simulations correspond to DMs of 40.7 (\MWI), 74.1 (\LMCI), and 125.9 (\LMCW) pc cm$^{-3}$, respectively, 
consistent with typical galactic values \citep{Crawford2001, Manchester2005, Prayag2026}. The \LMCW\ case shows the largest DM due to the presence of a ram-pressure-stripped tail of warm ionised gas extending through the MW’s CGM due to the LMC's motion, which increases the electron path length. Conversely, the \MWI\ case yields the lowest average DM, with values comparable to those seen for Galactic pulsars at $\sim 1$\,kpc distances, as verified with \textsc{PyGEDM} (\citealt{Price2021}, using pulsar data from \citealt{Cordes2002} and \citealt{Yao2017}).

By contrast, for $\alpha = 1/2$, $K$ reduces to $(L_p / f_c)^{1/2}$, the square root of the clumping-factor normalised path length. If we repeat our fitting procedure but fix $\alpha = 1/2$ and $\beta = 0$, we obtain best-fit values of $K = L_p/f_c = 1.7$ kpc, 5.5 kpc, and 15.8 kpc for the \MWI, \LMCI, and \LMCW\ cases, respectively. The \LMCW~case is again the largest, consistent with the presence of large amounts of extended extra-planar ionised gas due to ram pressure stripping of the simulated galaxy.

\subsection{What can we say about mass-weighted magnetic fields?}
\begin{figure*}
    \includegraphics[width=2\columnwidth]{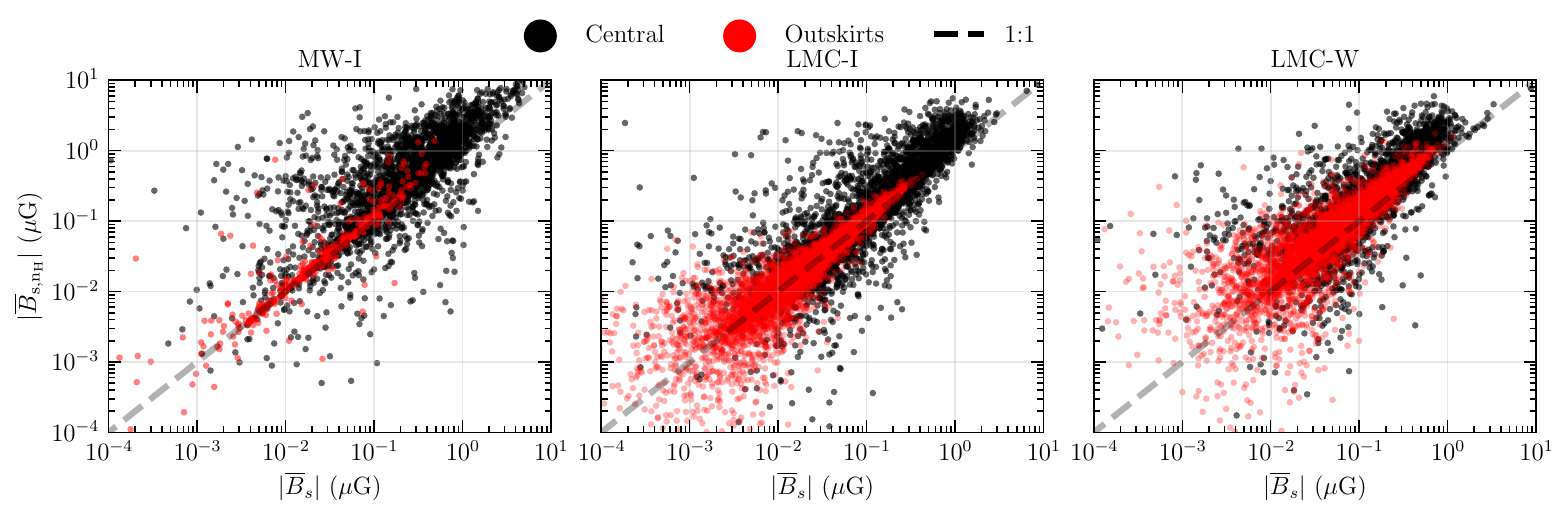} 
    \caption{Neutral hydrogen density (or mass) weighted magnetic field (\autoref{eq:BznH}) versus electron density weighted magnetic fields (\autoref{eq:Bzne}) for the three simulations (columns left to right). Black (red) points show values from central (outskirts) sightlines. The grey dashed lines show the one-to-one relation.\\}
    \label{fig:BVB_three_cases}
\end{figure*}

The results presented so far have focused on estimating the electron density-weighted magnetic fields, since RM only carries information about magnetic fields in the presence of free electrons. In this section, we check the usefulness of RM information in trying to predict mass-weighted magnetic fields instead, defined as
\begin{equation}
    \label{eq:BznH}
    \BznH \equiv \frac{\int \nH B_s\, \mathrm{d}s}{\int \nH \,\mathrm{d}s},
\end{equation}
where $\nH$ is the total number density of H nuclei, independent of their chemical state. \autoref{fig:BVB_three_cases} compares the mass-weighted magnetic fields ($\BznH$, y-axis) to the electron-weighted magnetic fields ($\Bz$, x-axis) we have considered thus far. From the Figure, it is clear that central galaxy sightlines tend to have systematically higher mean $\BznH$ than $\Bz$ and a large scatter in the relationship between the two, while in the outskirts, the relation between $\BznH$ and $\Bz$ more closely tracks the one-to-one line in terms of both mean and scatter. This difference can be understood as occurring because the central sightlines have both neutral and ionised gas in abundance, while in the outskirts most of the gas is ionised and thus $\BznH$ and $\Bz$ are nearly identical. The greatest deviation between $\BznH$ and $\Bz$ is in run \LMCW, where even the outskirts sightlines have a higher $\BznH$ than $\Bz$, which suggests that along those sightlines there is still a substantial amount of neutral gas that entraps magnetic fields. The neutral gas ram pressure stripped from the central ISM to the outskirts in the presence of the MW CGM wind can explain this effect.

\begin{table}
\centering
\footnotesize
\setlength{\tabcolsep}{3pt}
\caption{Same as \autoref{tab:missing_observations_free}, but for mass-weighted rather than $n_e$-weighted mean magnetic fields.}
\label{tab:Btrue_Bmeasured_nh_weighted}
\begin{tabular}{@{}c c c@{\hspace{0.2in}} c c c c@{}}
\hline
\multicolumn{3}{c}{Model} &
\multicolumn{4}{c}{Fit result} \\
Sim & Reg & Obs & $\alpha$ & $\beta$ & $\log K$ & $\sigma_{\chi}\,(\mathrm{dex})$ \\
\hline
\multirow{6}{*}{MW-I} & \multirow{3}{*}{C} & $N_\mathrm{\Hi, true}$, EM & 0.42 & -0.16 & 1.44 & 0.491 \\
 &  & $N_\mathrm{\Hi, obs}$, $\mathrm{EM_{ff}}$ & 0.41 & -0.15 & 1.46 & 0.493 \\
 &  & $N_\mathrm{\Hi, obs}$, $\mathrm{EM_{fb}}$ & 0.29 & -0.14 & 1.44 & 0.503 \\
\cline{2-7}
 & \multirow{3}{*}{O} & $N_\mathrm{\Hi, true}$, EM & 0.27 & 0.01 & 1.42 & 0.388 \\
 &  & $N_\mathrm{\Hi, obs}$, $\mathrm{EM_{ff}}$ & 0.27 & -0.00 & 1.43 & 0.402 \\
 &  & $N_\mathrm{\Hi, obs}$, $\mathrm{EM_{fb}}$ & 0.17 & -0.00 & 1.40 & 0.426 \\
\hline
\multirow{6}{*}{LMC-I} & \multirow{3}{*}{C} & $N_\mathrm{\Hi, true}$, EM & 0.53 & -0.42 & 1.74 & 0.363 \\
 &  & $N_\mathrm{\Hi, obs}$, $\mathrm{EM_{ff}}$ & 0.52 & -0.44 & 1.74 & 0.364 \\
 &  & $N_\mathrm{\Hi, obs}$, $\mathrm{EM_{fb}}$ & 0.47 & -0.49 & 1.79 & 0.367 \\
\cline{2-7}
 & \multirow{3}{*}{O} & $N_\mathrm{\Hi, true}$, EM & 0.30 & 0.03 & 1.77 & 0.213 \\
 &  & $N_\mathrm{\Hi, obs}$, $\mathrm{EM_{ff}}$ & 0.31 & 0.02 & 1.79 & 0.214 \\
 &  & $N_\mathrm{\Hi, obs}$, $\mathrm{EM_{fb}}$ & 0.27 & 0.03 & 1.81 & 0.214 \\
\hline
\multirow{6}{*}{LMC-W} & \multirow{3}{*}{C} & $N_\mathrm{\Hi, true}$, EM & 0.39 & -0.25 & 1.82 & 0.378 \\
 &  & $N_\mathrm{\Hi, obs}$, $\mathrm{EM_{ff}}$ & 0.39 & -0.25 & 1.82 & 0.375 \\
 &  & $N_\mathrm{\Hi, obs}$, $\mathrm{EM_{fb}}$ & 0.36 & -0.27 & 1.85 & 0.378 \\
\cline{2-7}
 & \multirow{3}{*}{O} & $N_\mathrm{\Hi, true}$, EM & 0.46 & -0.21 & 1.80 & 0.328 \\
 &  & $N_\mathrm{\Hi, obs}$, $\mathrm{EM_{ff}}$ & 0.38 & -0.16 & 1.82 & 0.332 \\
 &  & $N_\mathrm{\Hi, obs}$, $\mathrm{EM_{fb}}$ & 0.23 & -0.09 & 1.84 & 0.336 \\
\hline
\end{tabular}
\end{table}

We next repeat the exercise of fitting a model of the form given by \autoref{eq:logB_predict_conventional} to the $\BznH$ data, using exactly the same procedure as for $\Bz$ in \autoref{ssec:models}, with the exception that for brevity we only fit the case where $\alpha$, $\beta$, and $K$ are all left free. We report our best-fitting free parameter values in \autoref{tab:Btrue_Bmeasured_nh_weighted}. We find that the scatters in the fits $\sigma_\chi$ are consistently much larger than for fits to $\Bz$, particularly in galaxy centres. This is not surprising given that RMs do not directly constrain magnetic fields in neutral gas at all, but it does confirm that RM measurements offer only weak constraints on mass-weighted mean magnetic fields as opposed to electron density-weighted ones. Adding to that, RMs are useful for constraining mass-weighted mean fields only to the extent that the mass is primarily ionised, such as in the outskirts, where $\echi$ values are relatively smaller. 



\subsection{Comparison to the previous models}
\label{ssec:comparison}

In this section we compare our results to earlier models in the literature, which have used a variety of techniques to convert RMs to DMs and thence to magnetic field estimates. The first model to which we compare is case 1 from \citet{Kaczmarek2017}, which assumes that the neutral and ionised gas are well-mixed so that the depth of ionised gas $L_{\rm \Hii}$ can be expressed as a filling factor of ionised gas $f$ times the depth of neutral gas $L_{\rm \Hi}$. The DM in this case is then
\begin{equation}
    \label{DM:Kacz1}
    \mathrm{DM} = (\mathrm{EM} \ f \ L_{\rm \Hi})^{1/2}.
\end{equation}
They then assume fixed values $L_{\rm \Hi} = 5 \kpc$, and $f=0.5$. In our notation (\autoref{eq: DM_predict}), this model corresponds to $\alpha=0.5$, $\beta=0$, and $K = 2.5$ kpc.

Our second comparison model comes from \citet{Livingston2024}, though case 2 of \citet{Kaczmarek2017} is also very similar. This model assumes a constant ionisation fraction $X$ with well-mixed neutral and ionised gas phases, and a path length $L_{\rm \Hii} = f \ L_{\rm \Hi}$. In this case the DM can be expressed as
    \begin{equation}
        \mathrm{DM} = \frac{X \ \NH}{f \ L_{\rm \Hi}}.
\end{equation}
\citeauthor{Livingston2024} use the same $L_{\rm \Hi} = 5 \kpc$ as in the first model, but adopt $f = 1$ and consider a range of values for the ionisation fraction $X$; for our comparison purposes, we adopt $X = 0.2$, which is roughly the mean of the values they consider. In terms of our parameterisation, this model corresponds to adopting $\alpha = 0$, $\beta = 1$, and $K = X / f L_\mathrm{\Hi} = 0.04$ kpc$^{-1}$.

Our third comparison model is to adopt a single value of DM for the entire target galaxy, as done for example by \citet{Gaensler2005}. For our numerical comparison here we adopt the same DM as \citeauthor{Gaensler2005}: $57.4 \pc \cm^{-3}$. In our notation, this DM is the value of $K$, and obviously $\alpha = \beta = 0$.

The fourth model to which we compare is case 3 from \citet{Kaczmarek2017}, which assumes a fully ionised skin in thermal equilibrium with the neutral gas and density half that of the neutral gas $\langle \ne \rangle = \langle n_{\rm \Hi} \rangle /2$. The DM can then be expressed as
\begin{equation}
    \label{eq:DM_Kacz3}
    \mathrm{DM} = \frac{2 \ \mathrm{EM} \ L_{\rm \Hi}}{\NH},
\end{equation}
assuming the same $L_\mathrm{\Hi} = 5$ kpc as in the previous \citet{Kaczmarek2017} models. In our notation this corresponds to $\alpha = 1$, $\beta = -1$, and $K = 2 L_\mathrm{\Hi} = 10$ kpc.

For our fifth comparison case, we note that some studies have also assumed a model for the electron-density distribution across the galaxy to predict the DM -- for instance \citet{Shah2021} used an electron density model from \citet{Taylor1993} derived from pulsars in the Milky Way -- and assumed the same distribution for external spiral galaxies. For our test we use the same approach, omitting the terms in the \citeauthor{Taylor1993} model that correspond to spiral structure. In this case the model predicts an electron density
\begin{align}
    n_e(r,z) &= n_0 \operatorname{sech}^2\frac{r}{20\,\mathrm{kpc}} \operatorname{sech}^2\frac{z}{0.88\,\mathrm{kpc}} + {} 
    \notag \\
    & n_1 \exp\left[-\left(\frac{r-3.5\,\mathrm{kpc}}{1.8\,\mathrm{kpc}}\right)^2\right] \operatorname{sech}^2 \frac{z}{0.15\,\mathrm{kpc}}
\end{align}
where $n_0 = 0.0223$ cm$^{-3}$, $n_1 = 0.1$ cm$^{-3}$, and $(r,z)$ are the radial and vertical coordinates in a galactocentric cylindrical coordinate system. The DM along any sightline can then be obtained directly from integrating this density distribution, i.e., $\mathrm{DM} = \int n_e(r,z) \, \mathrm{d}z$.

\begin{figure*}
    \includegraphics[width=2\columnwidth]{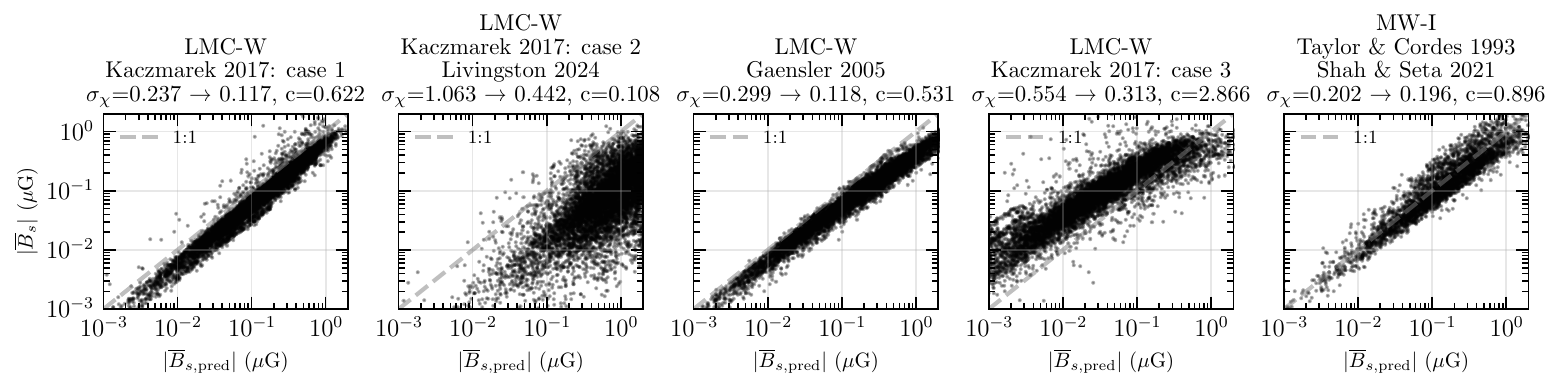} 
    \caption{True simulation magnetic fields (y-axis) versus magnetic fields predicted by five literature models (x-axis; outlined in \autoref{ssec:comparison}) when applied to our simulations. The title of each panel displays the unscaled $\echi$-error (left of arrow) and the best possible $\echi$-error for that model (right of arrow), which is obtained by scaling the predicted magnetic fields by a factor $c$.\\}
    \label{fig:compare_previous_models_LMC_MW_Bfield}
\end{figure*}

To test the performance of these five models, we use them to predict the DM along the same set of sightlines used in \autoref{sec:results}. We use only the \LMCW~run to test models one through four, and only the \MWI~data at $0^{\circ}$ inclination for the fifth model, since the first four models were developed specifically for the LMC and the final one assumes a spiral galaxy. In all cases we use the exact rather than observationally-inferred values of EM and $\NH$. We then predict the corresponding magnetic field as $\Bpredicted = \mathrm{RM}/(k_{\rm RM} \ \mathrm{DM})$. \autoref{fig:compare_previous_models_LMC_MW_Bfield} shows these predicted magnetic fields $\Bpredicted$ versus the true electron density-weighted magnetic field along each sighline $\Bz$. In this figure we show the scatter in dex $\sigma_\chi$ at the top of each panel.

Since these models have not been tuned to our simulations, some of the constant values used in them (e.g., the assumed value of $L_\mathrm{\Hi}$) may not be well-suited to our simulations, and thus using them might unfairly bias the comparison to our best-fit models. To mitigate this effect we also consider scaling the predicted magnetic fields by a constant value $c$, which we then optimise in order to minimise the scatter $\sigma_{\chi}$. We report both the resulting reduced $\sigma_{\chi}$ (values to the right of the arrow) and the corresponding scale factors $c$ in each panel as well.

For the \LMCW\ cases, we find that the first and the third models -- corresponding to assuming that DM $\propto$ EM$^{1/2}$ or simply using constant DM -- perform best, with minimum $\sigma_{\chi} \approx 0.12$ dex. The good performance from the third model is quite surprising, and may be due to DMs and EMs being spatially uncorrelated in the \LMCW\ case, a phenomenon that also explains whey this model has a lower value of $\alpha=0.22$ than the other two in our fits (see \autoref{tab:missing_observations_free}). Taking into account the value of the rescaling $c$, the model amounts to adopting a constant DM of $\sim108 \pc \cm^{-3}$ for the LMC. However, the value of $\sigma_\chi$ achieved by this model, and by the DM $\propto$ EM$^{1/2}$, remain noticeably larger than the values we can achieve leaving both $\alpha$ and $\beta$ free, or even simply leaving $\alpha$ free (cf.~\autoref{tab:missing_observations_beta0}).

Neither the second nor the fourth model -- assuming $\mathrm{DM}\propto \NH$ or $\mathrm{DM}\propto \mathrm{EM} / \NH$ -- produces accurate results. This failure occurs because the assumptions that lead to these models -- well-mixed neutral and ionised gas phases with a constant ionisation fraction for the second model, and the presence of ionised gas skins around the neutral gas for the fourth model -- are not satisfied in our simulations. The assumptions for the second model can be very weakly correct in some situations, e.g., the outskirts of galaxies, but even there, the best-fit value of the exponent $\beta$ on $\NH$ is much smaller than the value of unity adopted in this model. One is almost always better off using a constant DM value for the entire galaxy instead.

We demonstrate this point further in \aref{sec:robustness_tests}, where we show $\echi$ as a function of $\alpha$ and $\beta$ across the range of parameters that literature models have spanned. Our exploration there shows that $\echi$ worsens significantly as one moves away from the parameters listed in \autoref{tab:missing_observations_free}. This error topology explains the superior performance of models one and three relative to models two and four, as the former are more proximal to the best-fit global minima.

The fifth model, which predicts DMs by adopting $\ne$ from pulsar observations in the Milky Way, performs acceptably, $\sigma_{\chi} \approx 0.2$ dex. The marginal difference between the value of $\sigma_{\chi}$ we obtain without rescaling (i.e., with $c=1$) and with rescaling suggests that the \citet{Taylor1993} electron density distribution model is also a reasonably good description of our \MWI~simulation. Nonetheless, despite the reasonable performance of this model, the scatter is still noticeably worse than we can obtain using models derived from our numerical fits. We will use our models to provide our ``best practice'' recommendation to observers in the next section.

\subsection{Our final recommendations to observers}

\label{sec:to_observers}

We now come to our final recommendation for how observers should estimate DMs, and thus estimate magnetic fields from RM data. Our model predicting DMs is $\mathrm{DM} = K \ \mathrm{EM}^{\alpha} \ \NH^{\beta}$ (see \autoref{eq: DM_predict}), so an observer who wishes to use it must choose values for the three free parameters $\alpha$, $\beta$, and $K$. Our recommendations in this section therefore amount to strategies for choosing these values.

Since the recommended values of the free parameters differ slightly between central regions and the outskirts, a first step is identifying the region the observations are probing. One can identify the outskirts based on (in decreasing order of priority, based on available information)
\begin{enumerate}
    \item distance from the centre: if the deprojected radius is greater than $\sim 7.5 \kpc$ for a dwarf galaxy, or $\sim 16 \kpc$ for a MW mass galaxy \footnote{As it is tricky to obtain accurate deprojected radii for almost edge-on galaxies with high inclination angles, if the inclination angle $> 60^{\circ}$, it is acceptable to assume that the majority of sightlines probe outskirts (see \aref{sec:robustness_tests}).}
    \item neutral hydrogen density: if $\NH<2\times10^{19} \cm^{-2}$
    \item ionisation fraction: if more than $60\%$ gas is ionised across the majority of sightlines in a region.
\end{enumerate}

Once the observer has identified whether their sightlines are in the central regions or the outskirts, they can proceed to select the free parameters based on the galactic environment they are probing and the observations they have available, following the scenarios outlined below. Our best-fit numbers for various scenarios are provided in \autoref{tab:missing_observations_free}, \autoref{tab:missing_observations_alpha0}, \autoref{tab:missing_observations_beta0}, and \autoref{tab:missing_observations_alphabeta0}, but for convenience we have extracted the key numbers and summarised them in \autoref{tab:final_predictors}.

\begin{table*}
\centering
\begin{tabular}{l@{\hskip 1cm}ccc@{\hskip 1cm}ccc}
\hline\hline
Galaxy type & \multicolumn{3}{c}{Central regions\qquad\qquad\quad} & \multicolumn{3}{c}{Outskirts\qquad\qquad} \\
& $\alpha$ & $\beta$ & $\log K$ & 
$\alpha$ & $\beta$ & $\log K$
\\
\hline
\multicolumn{7}{c}{Scenario 1: EM estimator available} \\
\hline
Spiral &
0.34 & 0 & 1.64 &
0.44 & 0 & 1.65 \\
Isolated dwarf &
0.30 & 0 & 1.73 & 
0.39 & 0 & 1.80 \\
Interacting dwarf &
0.22 & 0 & 1.98 &
0.26 & 0 & 1.97 \\
\hline
\multicolumn{7}{c}{Scenario 2: EM estimator unavailable, \Hi~estimator available} \\
\hline
Spiral &
0 & 0.10 & 1.56 &
0 & 0.17 & 1.44 \\
Isolated dwarf &
0 & 0.01 & 1.87 & 
0 & 0.23 & 1.85 \\
Interacting dwarf &
0 & 0.12 & 2.03 &
0 & 0.16 & 1.98 \\
\hline
\multicolumn{7}{c}{Scenario 3: neither \Hi~nor EM estimators available} \\
\hline
Spiral &
0 & 0 & 1.61 &
0 & 0 & 1.29 \\
Isolated dwarf &
0 & 0 & 1.87 &
0 & 0 & 1.73 \\
Interacting dwarf &
0 & 0 & 2.10 &
0 & 0 & 1.98 \\
\hline
\end{tabular}
\caption{Final recommended parameters for dispersion measure estimation. Note that $K$ is in conventional units of pc$^{1-\alpha}$ cm$^{6\alpha+2\beta-3}$, so for scenario 3, $\alpha=\beta=0$, $K$ is simply the mean DM of the galaxy, in units of pc cm$^{-3}$. All the units and normalisations for different observables stated in this table are detailed in \autoref{eq:logB_predict_conventional}. The recommendations for Spiral, Isolated dwarf, and Interacting dwarf are based on simulations of an isolated MW-like galaxy, an isolated LMC-like galaxy, and the same LMC-like galaxy undergoing ram pressure stripping due to passage through the MW CGM, respectively.}
\label{tab:final_predictors}
\end{table*}

\begin{enumerate}
    \item \textbf{Scenario 1 - EMs available:} If an observer has access to observations of H$\alpha$ emission, free-free emission, or a similar tracer that allows direct estimation of the emission measure as a function of position, our experiments suggest that these provide the best available proxy for DM measurements, and that the observer should derive DMs from the EMs using the parameters provided in the first section of \autoref{tab:final_predictors}. Note that in this case we recommend that the observer not make use of $\NH$ values even if they are available, because they add no significant accuracy compared to relying on EMs alone. Based on our experiments, we expect an error of less than 0.1 dex when using the values from this model for magnetic field measurements. If both free-bound and free-free emission are available and equally well-measured, we recommend using free-free emission-derived EMs, but in practice the difference between the two is small, and thus the quality of the data available should be the primary deciding factor.

    \item \textbf{Scenario 2 - EMs unavailable but \Hi~available:} If no EMs are available, the second-best option is to use estimates of $\NH$ with the fit parameters provided in the second section of \autoref{tab:final_predictors}. This only offers a marginal improvement over simply assuming a constant DM value in the central regions of galaxies, but is helpful in the outskirts. The observer can expect magnetic field measurement errors at $\approx 0.2$ dex of accuracy when using this approach. 

    \item \textbf{Scenario 3 - No observational constraints on EM or \Hi~available:} 
    If both EM and \Hi~estimators are unavailable, we recommend that the observer simply adopt the mean $\log K$ values (which in this situation are equivalent to mean values of DM) that we find from our simulations, and summarise in the third section of \autoref{tab:final_predictors}. The observer can expect measurement errors of roughly 0.3 dex when using this methodology.
\end{enumerate}

Finally, it is worth mentioning one special case: if, on top of the scenarios above, one or more DM measurements are available, then we strongly recommend making use of those values to calibrate the values of $\log K$ rather than just relying on our recommendations based on simulations. In most such situations DMs are only available along a small number of sightlines (and if they are available on many, then the best strategy is simply to use them directly), and so an observer should use our suggested approaches to make DM predictions at the positions spatially the closest to the locations of the DM measurements, then adjust the value of $K$ to minimise the mismatch between the predicted and observed DMs. These values can then be substituted in any of the scenarios above for a more accurate model.

\subsection{Limitations and caveats}
We note here several caveats to our study. First, the isolated galaxy simulations -- \MWI\ and \LMCI\ -- were not designed to be accurate far from the galactic centre, and do not include a realistic circumgalactic medium component that would presumably become increasingly dominant at very large galactocentric radii. While a reasonable outskirts with high ionisation fractions forms in the \LMCI\ case, perhaps due to its smaller size, the disc in the \MWI\ simulation has a relatively sharp edge that leads to a poor sampling of the outskirts region. This means that our results for the outskirts in this simulation in particular should be treated with caution. By contrast, the \LMCW\ case was intentionally designed to simulate ram pressure stripping, and thus does produce a realistic environment around the galactic disc. In future work, we may improve our treatment of isolated galaxy simulations by adding a CGM component.
    
Second, we use the \gizmo\ code to simulate all our galaxies, which uses a divergence cleaning rather than a constrained transport method for MHD. It is therefore possible that some of our magnetic field structures may be affected by artificial numerical divergence. To date there have been only limited studies of RM statistics in MHD galaxy simulations \citep[e.g.,][]{Pakmor18a, Jung23a, Maconi25a}, and no exploration of the systematics between different numerical methods, let alone between different treatments of subgrid feedback. It is therefore unclear to what extent the results depend on such systematics. We strongly encourage other theorists to test our proposed scaling relations for their simulations, so a robust catalogue of scaling relations can be provided to aid observers in deriving accurate magnetic fields from RMs.
    
Lastly, in our radiative transfer calculations to obtain the mock observables that we use to feed our statistical analysis, we omit a number of real-world effects that can complicate relationships between these observables and the underlying quantities we use them to measure. In particular, we do not take into account the effects of dust and dust corrections in H$\alpha$ emission, and we do not consider the problem of differentiating free-free from synchrotron emission in radio observations. We also do not consider instrumental noise except in a very crude way by adopting a simple floor on observable surface brightnesses. In real observations with finite observational uncertainties, these measurement errors might dominate the uncertainties in magnetic field measurements, rather than the model calibration error on which we focus here. Consequently, the uncertainties we estimate in our final recommendation to observers (\autoref{sec:to_observers}) should be regarded as lower limits, and the strategy of which observable to use to estimate RMs and thus magnetic fields must be informed by the quality of observational data available as well as the theoretical accuracy with which a given observable can predict RMs.

\section{Conclusions}
\label{sec:conclusions}
In this study, we use MHD simulations of three galaxies -- an isolated Milky Way (MW)-like spiral, and isolated Large Magellanic Cloud (LMC)-like dwarf, and ram pressure stripped dwarf specifically designed to reproduce the interaction of the LMC with the MW's circumgalactic medium -- to explore methods for predicting dispersion measures (DMs), which are required in order to convert polarimetric measurements of the Faraday rotation measure (RM) into measurements of magnetic fields. Such RM-based measurements are vital because they are one of the few tools we have available to measure magnetic fields beyond the Milky Way. Next-generation radio facilities such as the SKA will provide large grids of them, but at present a limiting factor in exploiting such measurements is that they do not directly measure the magnetic field, only a combination of the magnetic field strength with the DM that characterises the total column density of electrons. It is generally not possible to measure DMs directly for extragalactic sources, and we must therefore rely on proxies of poorly-known accuracy to estimate them; characterising the quality of these proxies, and selecting the best strategies for using them, is the primary motivation for our work. To this end, we carry out radiative transfer post-processing on our simulations to create several mock observables -- emission measures (EMs) based on both free-free and free-bound emission, \Hi~column densities derived from 21 cm emission, and optical depths and spin temperatures derived from 21 cm absorption -- which we then compare statistically to the true DMs we can measure from the simulations.

We find that, within the limits of instrument sensitivity, EM is by far the most useful predictor of the DM out of the candidate proxies that we investigate, and that calibrations of DM from EM can achieve $\sim 0.1$ dex accuracy. In the absence of EM observations, $\NH$ is the second-best predictor of the DM, but is only accurate for observations targeting the predominantly ionised outskirts of galaxies rather than their neutral-dominated central regions; in the outskirts regions where $\NH$ is a useful proxy, it allows estimates of DM with $\sim 0.2$ dex scatter. We express both proxy estimators for DM with a simple analytic fitting formula $\mathrm{DM} = K \, \mathrm{EM}^{\alpha} \, \NH^{\beta}$, and provide estimates for the fitting constants $K$, $\alpha$, and $\beta$ suitable for use in a wide range of galactic environments. In our numerical tests, making use of the parameters we identify yields substantial improvements in the accuracy of recovered magnetic field strengths compared to existing calibration strategies in the literature. We summarise our final recommendations in \autoref{sec:to_observers} and \autoref{tab:final_predictors}.

Our numerical experiments also provide useful physical insight into \textit{why} some proxies perform better than others, and why the fitting coefficients $\alpha$ and $\beta$ assume the values that we find. The fitting coefficient $\alpha$, which calibrates inference of DMs from EMs, takes on values intermediate between 0 and 0.5. The former corresponds to a naive model in which all sightlines have the same total electron column density, and EMs vary only due to changes in how those electrons are arranged, and the latter to a naive model in which variations in EM from one sightline to another are driven entirely by differences in the total number of electrons with no change in spatial distribution. Real values are intermediate because both effects -- changes in the total electron column and the electron distribution -- contribute, with the amount of contribution from each channel varying slightly between inner and outer galaxies and between isolated and interacting systems, leading to slight changes in the best-fitting value of $\alpha$. The coefficient $\beta$ differs from zero only in outer galaxies where the gas is predominantly ionised, and trace amounts of \Hi~can act as proxies for the total, mostly-ionised electron column; in the inner parts of galaxies where the gas is mostly neutral, variations in the \Hi~column correlate little with variations in the ionised gas column and thus the DM. Even in outer galaxies $\beta \sim 0.2$ rather than $\sim 1$ as would be expected if \Hi~columns and ionised gas columns were simply proportional to one another; the smaller value reflects the fact that more gas is found in the neutral phase than in the ionised phase along lines of sight with higher total column densities, leading to a super-linear relationship between \Hi~and total column density and thus a sub-linear value of $\beta$.

We conclude with a consideration of future prospects. 
The next generation of polarimetric data coming from facilities such as the SKA will map the magnetised universe with unprecedented spatial resolution and sensitivity. In this new landscape, the challenge of understanding cosmic magnetism will no longer be limited by data availability, but by our ability to estimate magnetic fields from the available data as accurately as possible. The present study represents an initial step toward this broader goal, but is certainly not the final word on the matter. We encourage other theorists to undertake similar studies -- building comprehensive, physics-inspired parameter lists across diverse environments, objects, and numerical frameworks -- to refine magnetic field estimation techniques. Such efforts will be invaluable not only for advancing our understanding of cosmic magnetism but also for disentangling DM contributions from multiple components along sightlines, a frontier challenge that will become increasingly important as fast radio bursts (FRBs) join the toolkit of observers studying the ionised interstellar and intergalactic medium.

\section*{Acknowledgements}

We would like to thank Marijke Havekorn, Trey Wenger, Shane P. O’Sullivan, Alex Hill, Marc-Antoine Miville-Deschênes, Mordecai-Mark Mac Low, Shmuel Bialy, and Susan Clark for very insightful discussions during international conferences. We are grateful towards Amit Seta, Roland Crocker, Craig Anderson, Neco Kriel, Neelesh Amrutha, Taaseen Islam, Aditi Vijayan, Maja Jabłońska, Freeke van de Voort, Prachi Khatri, Thomas Rintoul, and other members from the groups of MRK and NMc-G for valuable discussions that greatly improved the quality of this study.

This research was partially funded by the Australian Government through Australian Research Council Australian Laureate Fellowships (project number FL220100020 awarded to MRK and project number FL210100039 awarded to NMc-G). ZH acknowledges support from NSFC through grant No. 12503026 and support from Boya Fellowship at Peking University. This research was undertaken with the assistance of resources from the National Computational Infrastructure (NCI Australia), an NCRIS enabled capability supported by the Australian Government, through award jh2.

\section*{Data Availability}
Simulation data related to this work is available at \url{https://datacommons.anu.edu.au/DataCommons/rest/display/anudc:6387}. Post-processed data related to this work will be shared upon reasonable request to the corresponding author.



\bibliographystyle{mnras} 
\bibliography{main} 




\appendix

\section{Robustness tests}
\label{sec:robustness_tests}
In this appendix, we check carry out a series of tests of the robustness of our results. 

First, in the main text, we make use of two simulations from \citetalias{Shah2025a}: referred to \LMCI~and \LMCW~here, corresponding to the cases \texttt{LMC-I-2-6-M} and \texttt{LMC-W-2-6-M} using the naming convention of \citetalias{Shah2025a}. To explore how our results depend on the simulation setup, we consider four more simulations from this paper, which we list in
\autoref{tab:free_parameter_robustness}. Each of these simulations targets an isolated LMC-like galaxy similar to \LMCI, but they differ in the initial conditions and resolution. The simulations use a four-character naming scheme, \texttt{I-B-b-r}, with \texttt{I} denoting an isolated simulation (all the simulations we test here are of type \texttt{I}), \texttt{B} and \texttt{b} denoting the initial large-scale organised and root mean square turbulent magnetic field strengths in $\muG$, respectively, and \texttt{r} indicating the resolution (\texttt{L} (low): 1000 $\Msun$, \texttt{M} (medium): 250 $\Msun$, \texttt{H} (high): 100 $\Msun$). We refer readers to \citetalias{Shah2025a} for full details on the simulation setups and how they differ. For each of the additional simulations, we repeat our analysis from the main text, for brevity focusing only on the $\alpha$, $\beta$ free case and using true EMs and $\NH$ values. \autoref{tab:free_parameter_robustness} shows the best-fit parameters and errors $\sigma_\chi$ we find. First we see that \texttt{I-0-12-M} and \texttt{I-2-6-M} yield results that are quite similar both to one another and to the case \LMCI~in the main text (c.f.~\autoref{tab:missing_observations_free}). This indicates that initial magnetic field strengths play only a marginal role in setting the values of the fit parameters. Comparing \texttt{I-2-6-L} and \texttt{I-2-6-H} to each other and the \LMCI~(which is identical to the other two in all aspects but resolution), we see that the value of $\alpha$ drops significantly when going from low resolution (\texttt{I-2-6-L}) to medium resolution (\texttt{I-2-6-M}, or \LMCI), but remains similar between medium and high resolution (\texttt{I-2-6-H}), indicating convergence in the values of free parameters in the medium resolution simulations from which we derive our results in the main text.

\begin{table*}
\centering
\caption{Best-fit parameters for the unconstrained (free $\alpha$, $\beta$, and $\log K$) model across various initial conditions. We use true EM and $\NH$ values to derive $\alpha$, $\beta$, $\log K$, and $\sigma_{\chi}$ for both the central regions and outskirts. Note that $K$ is in conventional units of pc$^{1-\alpha}$  cm$^{6\alpha+2\beta-3}$.}
\label{tab:free_parameter_robustness}
\begin{tabular}{l|cccc||cccc}
\hline
Simulation & \multicolumn{4}{c||}{Central} & \multicolumn{4}{c}{{Outskirts}} \\
\hline
 & $\alpha$ & $\beta$ & $\log_{10} K$ & $\sigma_{\chi}$ (dex) & {$\alpha$} & {$\beta$} & {$\log_{10} K$} & {$\sigma_{\chi}$ (dex)} \\
\hline
\texttt{I-0-12-M} & 0.29 & -0.05 & 1.76 & 0.065 & {0.35} & {0.02} & {1.80} & {0.015} \\
\texttt{I-2-6-M} & 0.34 & -0.09 & 1.77 & 0.058 & {0.41} & {-0.02} & {1.79} & {0.012} \\
\texttt{I-2-6-L} & 0.46 & -0.09 & 1.72 & 0.029 & {0.38} & {-0.01} & {1.77} & {0.012} \\
\texttt{I-2-6-H} & 0.32 & -0.07 & 1.75 & 0.074 & {0.37} & {0.01} & {1.81} & {0.012} \\
\hline
\end{tabular}
\end{table*}

\begin{figure*}
    \includegraphics[width=2\columnwidth]{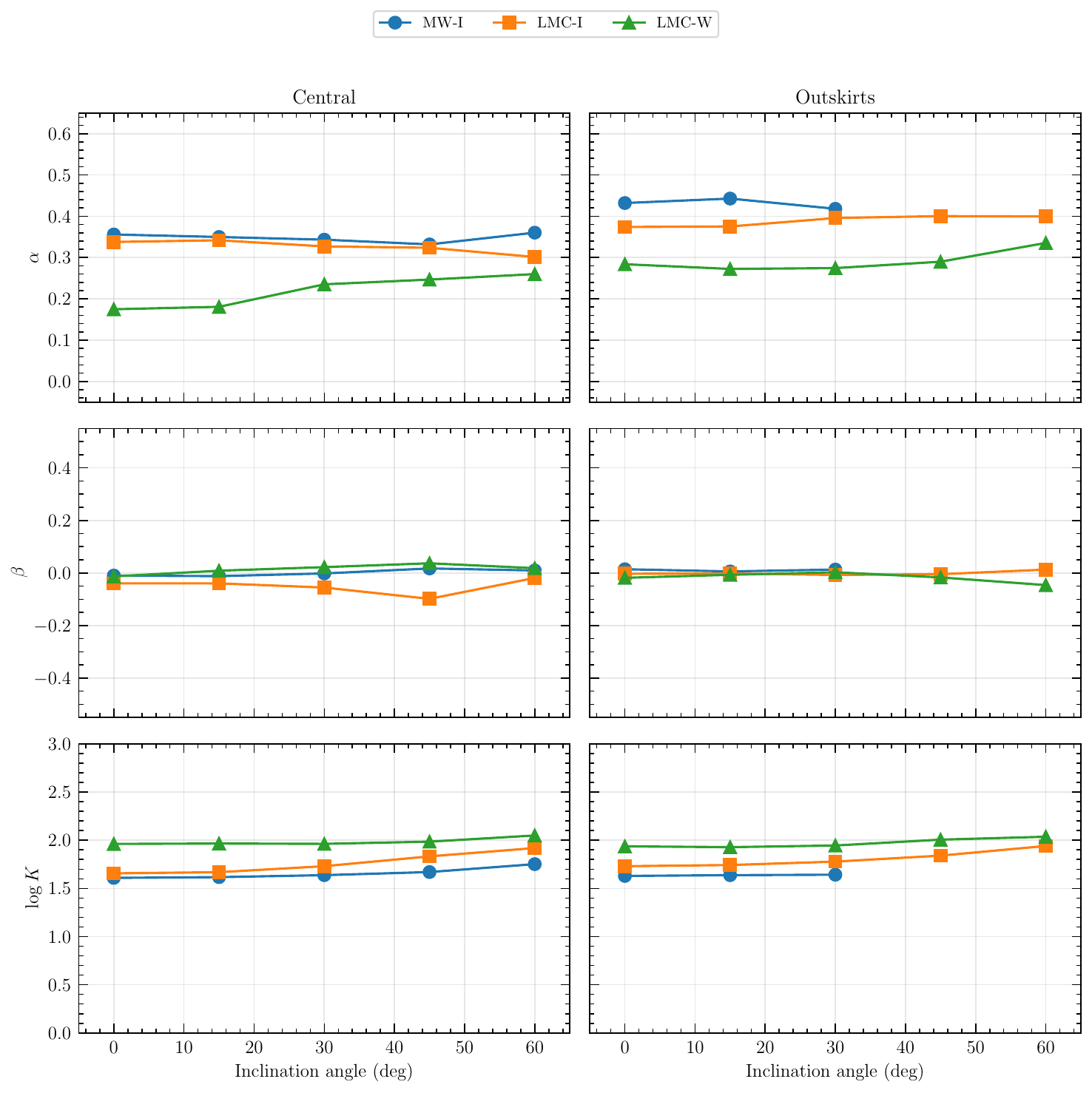} 
    \caption{Free parameters $\alpha$ (top row), $\beta$ (middle row), and $\log K$ (bottom row) plotted as a function of observed inclination angle for various galaxy cases. The left and right columns display results for the central regions and outskirts, respectively. We skip the fitting procedure whenever we obtain $<20$ sightlines, applicable only for some inclination angles in the \texttt{MW-I} case. Note that $K$ is in conventional units of pc$^{1-\alpha}$  cm$^{6\alpha+2\beta-3}$. \\}
    \label{fig:alpha_beta_inclination}
\end{figure*}

Second, we examine the effects of inclination angle by repeating the analysis presented in the main text at inclinations from $0^\circ - 60^\circ$, corresponding to face-on to almost edge-on orientations. \autoref{fig:alpha_beta_inclination} shows how the best-fit parameters vary as a function of the inclination angle for both central regions and outskirts. The variation of all the free parameters is minimal with varying inclination angles, with all parameters maintaining values very similar to the ones shown in \autoref{tab:missing_observations_free} of the main text. Thus, we conclude that the inclination angle does not have a major effect on the choice of magnetic field prediction models, at least for moderate inclinations of up to $60^\circ$.

\begin{figure*}
    \includegraphics[width=2\columnwidth]{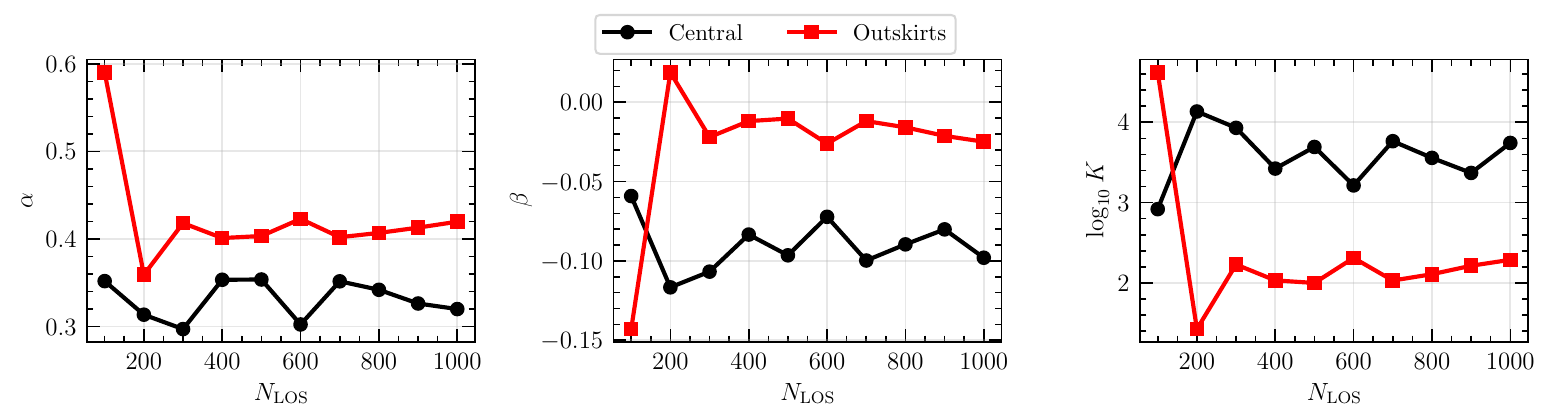} 
    \caption{Best-fit free parameters $\alpha$, $\beta$, and $\log K$ as a function of the number of sightlines available for the fitting procedure. The sightlines are picked randomly from a spatially uniform distribution. We use \LMCI\ at $34.7^{\circ}$ inclination angle as a representative simulation for this (other cases give similar results). Note that $K$ is in conventional units of pc$^{1-\alpha}$. \\}
    \label{fig:plot_sampling_robustness_I26M}
\end{figure*}

Next, we investigate whether our samples of sightlines are large enough for our fitting parameters to have converged. To perform this experiment, we repeat our analysis of the \LMCI\ simulation for varying numbers $N_\mathrm{LOS}$ of sample sightlines. For this test we use the model with $\alpha$ and $\beta$ both free, and use the same inclination angle $34.7^\circ$ as in the main text. We show the fit parameters as a function of $N_\mathrm{LOS}$ in \autoref{fig:plot_sampling_robustness_I26M}. It is clear from the figure that we can obtain stable values of the free parameters when we have $> 300$ sightlines for both central and outskirts; we ensure that this condition is satisfied for all the results in our main text. 

\begin{figure*}
    \includegraphics[width=2\columnwidth]{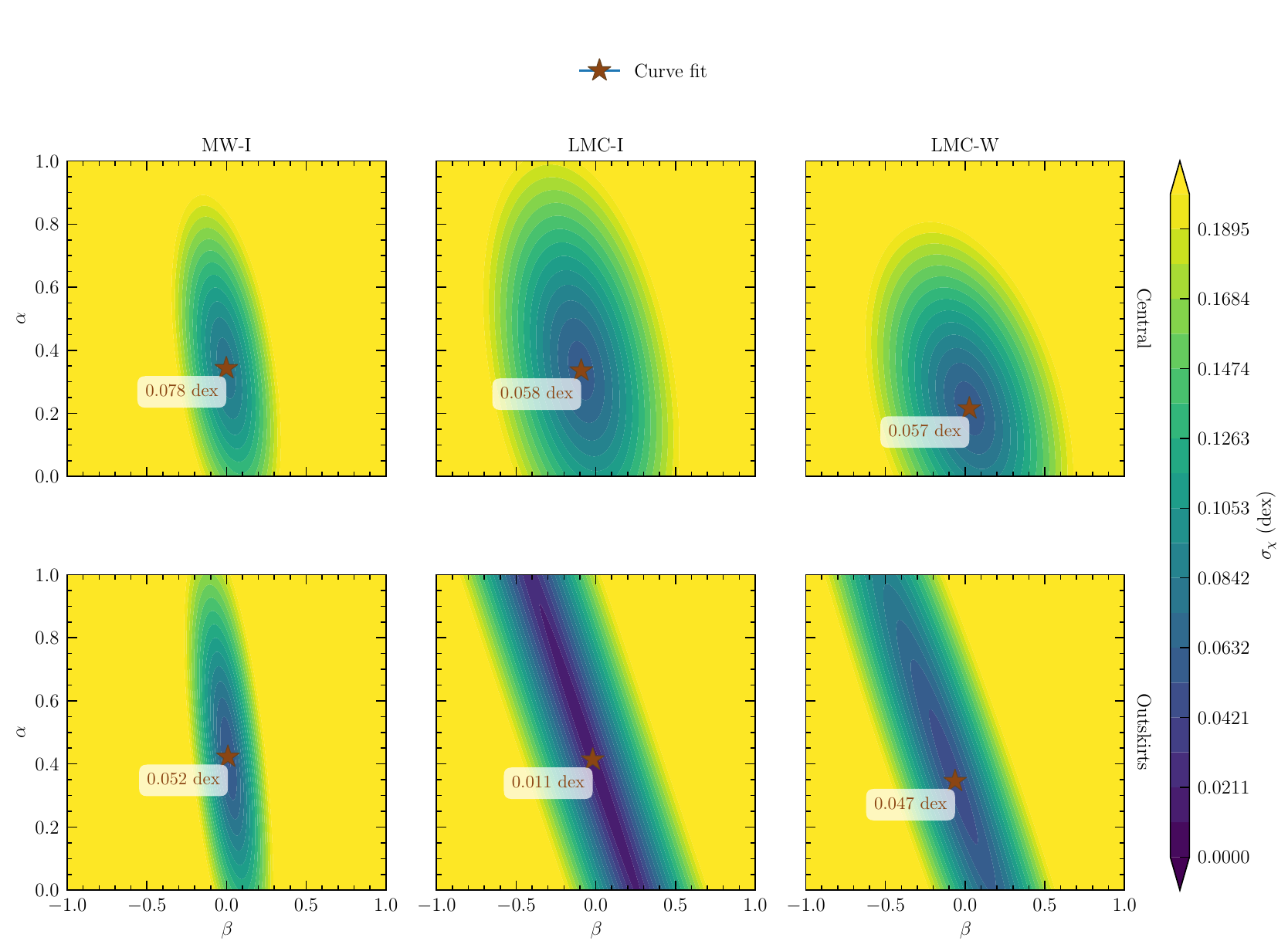} 
    \caption{Minimum $\echi$ values (in dex) as a function of $\alpha$ and $\beta$ for our simulated galaxies. Columns represent different simulations, and rows denote the central versus the outskirts regions. The parameter space spans $\alpha \in [0, 1]$ and $\beta \in [-1, 1]$, ensuring that we cover all the models discussed in \autoref{ssec:comparison}. Star markers indicate the $\alpha$ and $\beta$ values derived from our curve fits, and the corresponding annotations indicate the global minimum of $\echi$ achieved at these points. \\}
    \label{fig:rms_err_surface_central_outskirts}
\end{figure*}

Finally, in \autoref{fig:rms_err_surface_central_outskirts} we show the scatter $\echi$ across a broad range of $\alpha$ (0 to 1) and $\beta$ (-1 to 1) values, in order to demonstrate both that our fits represent a unique minimum in this space, and to quantify the relatively worse performance of earlier literature models -- the range over which we plot is selected specifically to cover the parameter space of all previously used models in the literature (see \autoref{ssec:comparison}). Since our parameter space is three-dimensional, including $\alpha$, $\beta$, and $\log K$, to reduce the dimensionality from 3D to 2D and aid in visualisation, at each $\alpha$ and $\beta$ we show the minimum value of $\echi$, corresponding to a value of $\log K$ (denoted as $\log K_{\rm best}$) obtained by differentiating $\echi (\alpha_i, \beta_i, \log K)$ with respect to $\log K$ and solving for the value of $\log K$ that makes the derivative zero. Mathematically, this condition can be expressed as
\begin{equation}
    \frac{d\echi (\alpha_i, \beta_i, \log K)}{d\log K} = 0,
\end{equation}
which has the solution\footnote{For the purposes of this calculation we use EM$_\mathrm{true}$ and $N_\mathrm{HI,true}$, but the results using $\mathrm{EM_{fb, obs}}$, $N_\mathrm{\Hi,obs}$ and $\mathrm{EM_{ff, obs}}$, $N_\mathrm{\Hi,obs}$ values are qualitatively similar.}
\begin{align}
    \label{eq: logK_best}
    \log \left( \frac{K_{\rm best}}{\cm^{5\alpha_i + 2\beta_i - 2}} \right)
    &=  \frac{1}{N_{\rm LOS}} \sum_j \biggl[
        \log\left( \frac{|{\rm RM}|}{\mathrm{rad\,cm^{-2}}} \right)_j \\
    &\quad
        - \log\left( \frac{K_{\rm RM}}{\mathrm{rad\,G^{-1}}} \right)
     \notag  - \alpha_i \log\left( \frac{{\rm EM}}{\mathrm{cm^{-5}}} \right)_j \notag \\
    &\quad
    - \beta_i \log\left( \frac{N_{\rm H}}{\mathrm{cm^{-2}}} \right)_j  -\log\left(\frac{|\Btrue|}{\mathrm{G}}\right)_j \biggr].
\end{align}

The plots of $\echi$ reveal several key features. There exists a single global minimum for the best-fitting free parameters, indicating that convergence to local minima is not a concern. The lowest dex contour (with a range of approximately $\sim 0.01$ dex) in each panel spans a relatively broad region of $\alpha$ and $\beta$. For central regions, this range is approximately $\pm 0.05$ around the best-fitting $\alpha$ value and $\pm 0.01$ around the $\beta$ value. In particular, the lowest dex contour consistently passes through $\beta=0$, highlighting that $\NH$ measurements are redundant when EM data are available. The width of the low-error contours in $\alpha$ further suggests that precise tuning of $\alpha$ and $\beta$ is not critical for constructing accurate magnetic field models; moderate deviations are tolerable. However, literature models that deviate very substantially from the best-fit -- for example those with $\beta = 1$ or $-1$, perform much worse.

\section{Numerical method for evaluating line of sight integrals}

\label{sec:AdaptiveBinning}

All the observable quantities defined in \autoref{sec:MockObservables} are derived by evaluating integrals along a line of sight. In this appendix, we describe the algorithm by which we evaluate such LOS integrals through a simulation domain. This requires some care, because the simulations we are processing were carried out using the Lagrangian \gizmo~code, which describes the continuous medium through which our LOS passes in terms of the properties of a series of fluid parcels located at arbitrary positions, and with different effective volumes. In particular, the raw \gizmo~output from which we begin provides the position $\mathbf{x}_i$ and local smoothing length $h_i$ at the position of each particle, along with a variety of other intensive quantities such as gas temperature and chemical composition. The smoothing length in turn is a parameter for the standard cubic spline kernel \citep{Hopkins2015std} defining the contribution of each particle's properties to the fluid field around it:
\begin{equation}
W(u, h) = \frac{8}{\pi h^3}
\begin{cases}
1 - 6u^2 + 6u^3, & 0 \leq u < 0.5 \\
2(1-u)^3, & 0.5 \leq u < 1 \\
0, & u \geq 1
\end{cases}
\end{equation}
where, $u = |\mathbf{x} - \mathbf{x}_i|/h_i$, $\mathbf{x}$ is an arbitrary position in the domain, $\mathbf{x}_i$ is the position of particle $i$, and $h$ is the smoothing length. We calculate the value of any physical quantity $Q$ at an arbitrary location $\mathbf{x}$ with a kernel function weighted average; i.e., $Q = \sum_{i} Q_i\, W(u_{i, n}, h_i)/\sum_{i} W(u_{i, n}, h_i)$, where $Q_i$ is the value of that quantity for particle $i$.

\begin{figure*}
    \includegraphics[width=2\columnwidth]{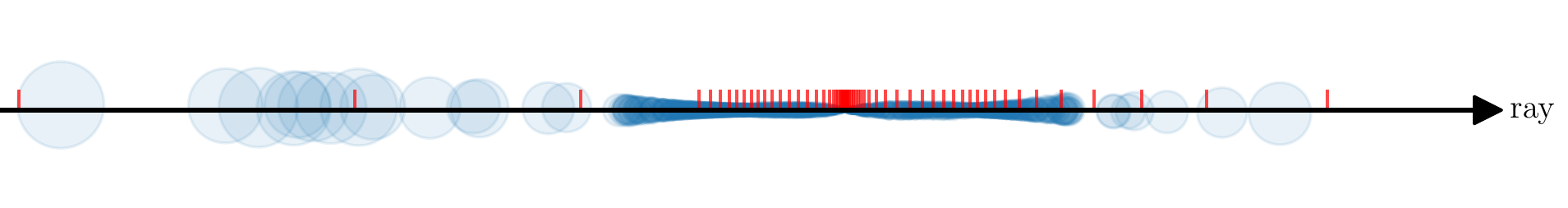} 
    \caption{An example sightline with gas particles (blue circles with radii proportional to smoothing lengths and a cutoff on the highest radius) and the location of the selected bin edges (red vertical lines). There is a much higher concentration of bins close to the centre of the ray, where the galaxy centre resides. Since the gas particles there have small smoothing lengths, we ensure an efficient allocation of computational resources. \\}
    \label{fig:rayTrace}
\end{figure*}

The final part of our algorithm is a method for choosing positions $\mathbf{x}$ along each LOS at which to evaluate the gas positions, and thereby to construct a discrete approximation to the integrals appearing in the definitions of our various quantities. This requires some care, because there is a very wide range of smoothing lengths, and thus brute force algorithms (e.g., choosing points separated by a distance smaller than the smallest smoothing length for any particle) are impractically expensive. Instead, we apply an adaptive algorithm, whose steps are as follows:
\begin{enumerate}
    \item First, we rotate the galaxy according to the desired inclination angle, such that the observer is always located along the $+z$ direction. For each ray passing through a point $\{x_j, y_j\}$ in the plane of the sky, we select all gas particles that satisfy $u_i = [(x_i - x_j)^2 + (y_i - y_j)^2]^{1/2}/h_i \leq 1$,
    where $i$ indexes the gas particles, $j$ specifies the coordinates through which the ray passes, and $h_i$ is the kernel length of particle $i$. This criterion includes all particles whose kernel overlaps the given sightline; we will only use these particles in the subsequent steps.
    
    \item Next, we start the adaptive binning process. Assuming a spherical kernel around particles, we identify all locations where particle kernels intersect with the ray (say $\{z_{\rm intersect}\}$); the start of the ray is the minimum $z$ of those intersections, and the end of the ray is the maximum $z$. Let these points be $z_{0}$ and $z_{-1}$, respectively. The length of the next step from $z_0$ is decided based on the effective kernel length $h_{\rm eff}$ at that location $\{x_j, y_j, z_0\}$, given by $h_{\rm eff, 0} = \sum_{i} h_i\, W(u_{i, 0}, h_i)/\sum_{i} W(u_{i, 0}, h_i)$. Thus, the position of the next bin edge is $z_1 = z_0 + K_{\rm adapt} \times h_{\rm eff, 0}$, where $K_{\rm adapt}$ is a dimensionless tuning parameter. The general expression for this process is $z_n = z_{n-1} + K_{\rm adapt} \times h_{\rm eff, n-1}$, where $n$ is steps along the ray. This process is repeated until $z_n \geq z_{-1}$. In the event where $\sum_{i} W(u_{i, n}, h_i)=0$; i.e., there are no contributing particles at a particular $z_n$, then the next step ensures an overlap with the next closest particle kernel: $z_{n+1} = \min(\{z_{\rm intersect}\}\geq z_n)$.

    \item The parameter $K_{\rm adapt}$ tunes the step sizes along the ray, with smaller values corresponding to finer binning. We begin with $K_{\rm adapt}=1$, generate the ray, and compute the mock observables. We calculate the physical quantities $Q_n$ at an arbitrary location $\{x_i, y_i, z_n\}$. For convergence testing, we use a set of representative observables -- 21 cm optical depth $\tau$, emission measure (EM), and brightness temperature $T_{\rm b}$ -- chosen to ensure convergence is achieved for a variety of independent physical properties. We then reduce $K_{\rm adapt}$ by half and recompute the observables. This process is repeated until the average percentage change between successive iterations for all representative observables is less than $1\%$. Such an adaptive binning approach guarantees that mock observables are calculated with sufficient accuracy while maintaining computational efficiency.
    
\end{enumerate}
\autoref{fig:rayTrace} shows a low-resolution working example of the algorithm, displaying the $z$-positions of gas particles (blue spheres with radii proportional to kernel lengths) and the adaptive steps (red vertical lines). The galaxy lies at the ray centre, where particles are denser and have smaller kernels. The algorithm automatically takes finer steps near the galaxy centre and coarser steps farther away. 

\bsp	
\label{lastpage}
\end{document}